\documentclass[
    reprint,
    superscriptaddress,
    amsmath, amssymb,
    aps,
    prx,
    floatfix,
]{revtex4-2}

\usepackage{newtxtext, newtxmath}

\usepackage{color, xcolor}
    \definecolor{mygreen}{rgb}{0,0.6,0}
    \definecolor{mygray}{rgb}{0.5,0.5,0.5}
    \definecolor{mymauve}{rgb}{0.58,0,0.82}
    \definecolor{myyellow}{rgb}{0.9, 1, 1}
    \definecolor{myblue1}{RGB}{64,57,144}
    \definecolor{myblue2}{RGB}{128,166,226}
    \definecolor{myblue3}{RGB}{70,217,255}
    \definecolor{myyellow}{RGB}{251,221,133}
    \definecolor{myred1}{RGB}{207,67,62}
    \definecolor{myred2}{RGB}{244,111,67}

\usepackage{amsmath, mathtools, commath, bm, amssymb}
\usepackage{framed}
\usepackage{extarrows}
\usepackage{booktabs}

\usepackage{graphicx}
\usepackage{geometry}
\usepackage{graphics}
    \graphicspath{{./figures/}}
\usepackage{setspace}
\usepackage[unicode=true, bookmarks=true, bookmarksnumbered=false, bookmarksopen=false, breaklinks=false, pdfborder={0 0 1}, backref=false, colorlinks=false, hidelinks]{hyperref}
    \hypersetup{linkcolor=black, urlcolor=black, citecolor=black, pdfstartview={FitH}}

\DeclareSymbolFont{CMlargesymbols}{OMX}{cmex}{m}{n}
\let\sumop\relax\let\prodop\relax
\DeclareMathSymbol{\sumop}{\mathop}{CMlargesymbols}{"50}
\DeclareMathSymbol{\prodop}{\mathop}{CMlargesymbols}{"51}

\SetSymbolFont{operators}{normal}{OT1}{ntxtlf}{m}{n}
\SetSymbolFont{operators}{bold}{OT1}{ntxtlf}{b}{n}

\usepackage[braket, qm]{qcircuit}

\newcommand{\on}[1]{\bm{#1}}

\renewcommand{\vec}[1]{{\bm{#1}}}

\renewcommand{\ge}{\geqslant}
\renewcommand{\le}{\leqslant}
\newcommand{\ee}{\mathrm{e}}

\newcommand{\wb}[1]{\mskip.5\thinmuskip\overline{\mskip-.5\thinmuskip {#1} \mskip-.5\thinmuskip}\mskip.5\thinmuskip}

\renewcommand{\ket}[1]{|#1\rangle}
\renewcommand{\bra}[1]{\langle#1|}

\newcommand{\beginsupplement}{%
	\setcounter{table}{0}
	\renewcommand{\thetable}{S\arabic{table}}%
	\setcounter{figure}{0}
	\renewcommand{\thefigure}{S\arabic{figure}}%
	\setcounter{equation}{0}
	\renewcommand{\theequation}{S\arabic{equation}}%
	\setcounter{section}{0}
	\renewcommand{\thesection}{\arabic{section}}%
}

\usepackage{nomencl}
\makenomenclature

\begin{document}

\title{Quantum lattice Boltzmann method for simulating nonlinear fluid dynamics}

\author{Boyuan Wang}
\affiliation{%
 State Key Laboratory for Turbulence and Complex Systems, College of Engineering, Peking University, Beijing 100871, PR China}
\author{Zhaoyuan Meng}
\affiliation{%
 State Key Laboratory for Turbulence and Complex Systems, College of Engineering, Peking University, Beijing 100871, PR China}
\author{Yaomin Zhao}
\affiliation{%
 State Key Laboratory for Turbulence and Complex Systems, College of Engineering, Peking University, Beijing 100871, PR China}
\affiliation{HEDPS-CAPT, Peking University, Beijing 100871, PR China}
\author{Yue Yang}%
\email{yyg@pku.edu.cn}
\affiliation{%
 State Key Laboratory for Turbulence and Complex Systems, College of Engineering, Peking University, Beijing 100871, PR China}
\affiliation{HEDPS-CAPT, Peking University, Beijing 100871, PR China}

\begin{abstract} 
Quantum computing holds great promise to accelerate scientific computations in fluid dynamics and other classical physical systems. While various quantum algorithms have been proposed for linear flows, developing quantum algorithms for nonlinear problems remains a significant challenge. 
We introduce a novel node-level ensemble description of lattice gas for simulating nonlinear fluid dynamics on a quantum computer. This approach combines the advantages of the lattice Boltzmann method, which offers low-dimensional representation, and lattice gas cellular automata, which provide linear collision treatment. Building on this framework, we propose a quantum lattice Boltzmann method that relies on linear operations with medium dimensionality. 
We validated the algorithm through comprehensive simulations of benchmark cases, including vortex-pair merging and decaying turbulence on $2048^2$ computational grid points. The results demonstrate remarkable agreement with direct numerical simulation, effectively capturing the essential nonlinear mechanisms of fluid dynamics. This work offers valuable insights into developing quantum algorithms for other nonlinear problems, and potentially advances the application of quantum computing across various transport phenomena in engineering. 
\end{abstract}
\maketitle

\let\oldaddcontentsline\addcontentsline
\renewcommand{\addcontentsline}[3]{}

\noindent Classical numerical methods for simulating fluid dynamics and general nonlinear problems demand substantial computational resources. For instance, the direct numerical simulation (DNS) of turbulence faces significant challenges due to the cubic growth of computational cost with the Reynolds number ($\mathrm{Re}$), which is necessary for resolving multiscale flow structures \cite{ishihara2009study, Shen2024_Designing}. Quantum computation, leveraging the principles of superposition and entanglement inherent in quantum states, has shown an exponential advantage over classical computational methods in certain specific problems \cite{harrow2009quantum,shor1994algorithms}. 
Consequently, quantum computing for fluid dynamics (QCFD) \cite{fukagatatowards2022,Bharadwaj2023_Hybrid,Liu2021_Efficient,Meng2023_Quantum,Kumar2024_Decomposition,Meng2024_Simulating,Itani2022_Analysis, Gourianov2022_A} is anticipated to tackle turbulence, one of most challenging problems in classical physics \cite{Feynman2015_The} and nonlinear dynamics.   

Classical and quantum systems exhibit intrinsic differences: fluid dynamics is nonlinear and irreversible, whereas quantum computing relies on linear and unitary operations. This fundamental disparity presents a substantial challenge in simulating nonlinear dynamics using quantum computers \cite{succiquantum2023,tenniequantum2025}. 
Several approaches have been proposed to partially address this challenge.  
The Navier-Stokes (NS) equations governing fluid motion have been transformed into the Schrödinger equation and solved via Hamiltonian simulation \cite{Meng2023_Quantum,Meng2024_Quantum,Meng2024_Simulating}. They have also been discretized into a set of nonlinear ordinary differential equations and then linearized \cite{gaitan_finding_2020,Liu2021_Efficient}. Additionally, quantum-classical hybrid methods have been employed to solve these equations \cite{Chen2024_Enabling, Lapworth2022_A, Pfeffer2022_Hybrid, Bharadwaj2023_Hybrid}. 
Moreover, the general partial differential equations can be transformed into linear ones by increasing the dimensionality of the problem \cite{Jin2023_Time,Tennie2024_Solving,Joseph2020_Koopman,Lu2024_Quantum}. 

In particular, the earliest attempt \cite{yepez2001quantum} in QCFD used the lattice Boltzmann method (LBM) \cite{kruger_lattice_2017,chen1998lattice}, which provides a natural framework for addressing nonlinearity in fluid dynamics \cite{qianLatticeBGKModels1992}. By reformulating the NS equations into a higher-dimensional lattice gas representation, LBM inherently transforms the nonlinear convection term in physical space into a linear transport term in phase space. However, the collision term in LBM, usually treated by the Bhatnagar-Gross-Krook (BGK) model \cite{qianLatticeBGKModels1992}, introduces additional nonlinearity, posing a significant challenge for its quantum implementation. 

Therefore, developing a linear yet physically accurate collision model is crucial for the quantum lattice Boltzmann method (QLBM) \cite{Itani2022_Analysis,budinskiquantum2021,Kumar2024_Decomposition}. In existing studies, the nonlinear part of the collision term has often been neglected to simulate simple convection-diffusion problems \cite{wawrzyniakquantum2025,Budinski2022_Quantum,Kocherla2024_Fully,dinesh2024quantum,kocherla2024two,bediche_full_2024}. The nonlinearity in the collision term has been eliminated using the Carleman linearization, but this approach results in an exponential increase in the dimension of the linearized system with scale \cite{Itani2024_Quantum,Sanavio2024_Three,sanaviocarlemanlatticeboltzmann2025,Itani2022_Analysis}. Another collision model in the lattice-gas cellular automata (LGCA) \cite{HPP1973} is linear and can be integrated into QLBM \cite{shakeel2013quantum,singh_emergence_2024}, but its quantum algorithms suffer from excessively high dimensionality \cite{wolf-gladrowLatticeGasCellular2000}. Consequently, an efficient QLBM algorithm for simulating nonlinear fluid dynamics has yet to be developed. 
    
Here we introduce a QLBM algorithm with a novel ensemble description of lattice gas. This ensemble assumes that each computational node contains multiple realizations, with all nodes being mutually independent. Our QLBM combines the strengths of both the LBM and LGCA, featuring medium dimensionality and linear collision treatment.  
In the implementation, the probability distribution of the ensemble is encoded into amplitudes of quantum states. The evolution of the quantum states is driven by linear transport and collision steps. Additionally, an H-step is employed to keep the ensemble near the equilibrium state. This involves disentangling qubits and increasing quantum entropy. The flow field is measured only at the end of simulation, thereby avoiding the need for frequent intermediate measurements and state preparations.

Notably, our QLBM is validated through simulations of vortex pair merging and turbulence on $2048^2$ computational grid points, representing the largest-scale case in QCFD to date. 
For comparison, the size of most existing QCFD simulations is limited to the order of $10^3$ to $10^4$ grid points. 
This successful application demonstrates the capability of QCFD to address highly nonlinear flow prediction, and provide insights into developing quantum algorithms for other nonlinear problems in various engineering applications. 

\begin{figure*} 
    \centering
    \includegraphics[scale=1]{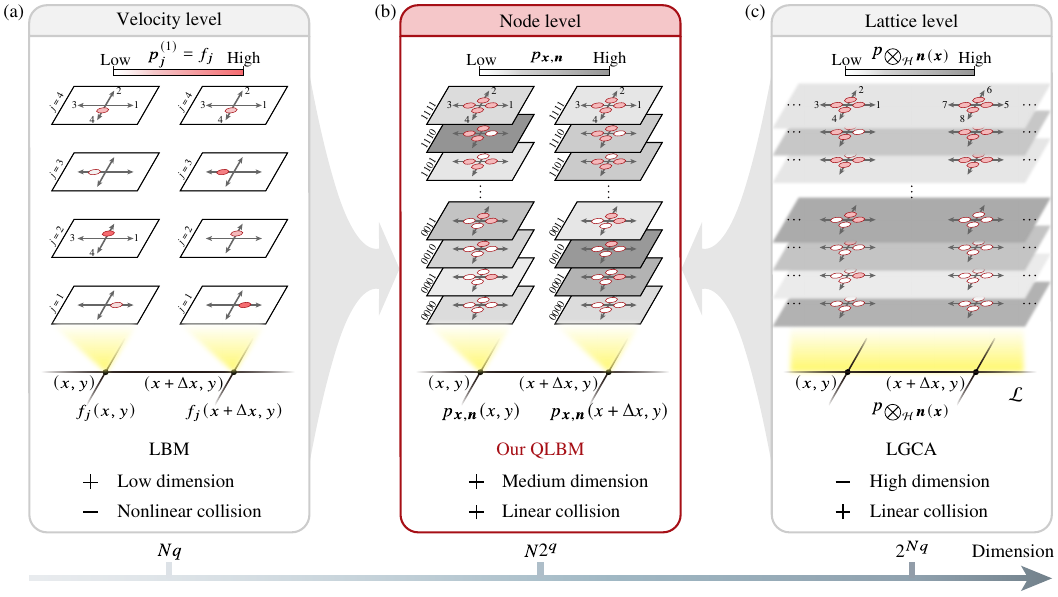}
    \caption{Schematic of three-level ensemble descriptions for lattice gas: (a) velocity-level ensemble, corresponding classical LBM; (b) node-level ensemble, corresponding to our QLBM; (c) lattice-level ensemble, corresponding to LGCA.
    The lattice $\mathcal{L}$ comprises discrete nodes (black dots). Each node represents a local flow state, and it consists of $q = 4$ cells (red circles) for the D2Q4 model. 
    Each card (containing nodes) represents a realization. 
    The probability of each realization is depicted by the saturation level of circles in (a) and the saturation level of cards in (b) and (c). 
    Our QLBM is featured by the medium dimensionality and linear collision treatment, suitable for an efficient quantum algorithm. 
    The particle collision in LBM in (a) involves nonlinear calculations with probability values for different cells, whereas the collision in our QLBM in (b) or LGCA in (c) is characterized by a linear combination of different realizations.} 
    \label{fig:QLBM_levels}
\end{figure*}

\section*{Results}
\noindent\textbf{Method overview}

\noindent The QLBM is used to solve the Boltzmann equation, which describes the evolution of the statistical distribution function of a particle system, and reduces to the NS equations at a small Knudsen number.
In classical LBM or LGCA, the Boltzmann equation is solved using lattice gas models by discretizing the particle velocity into $q$ discrete values through various DdQq models \cite{chen1998lattice} within a node in a $d$-dimensional space. 
As illustrated in Fig.~\ref{fig:QLBM_levels}, the lattice $\mathcal{L}$ comprises $N$ discrete nodes (black dots). 
Each node represents a local flow state at a spatial location $\vec{x} = (x, y)$ and stores particle properties.  
The particles transport from one node to another along $q$ predefined discrete directions (arrows), and they collide at the nodes. 
The collision of particles within a node drives the local lattice gas toward an equilibrium state. 
Each node consists of $q$ cells (red circles), and each cell can be either empty (open circles) or occupied (closed circles) by at most one particle with a specified velocity.

As illustrated in Figs.~\ref{fig:QLBM_levels}(a) and (c), neither classical LBM nor LGCA can simultaneously achieve manageable dimensionality and involve linear particle collisions. To address this, we propose a meso-level ensemble, situated between the LBM and LGCA, as shown in Fig.~\ref{fig:QLBM_levels}(b). This ensemble forms the basis of our QLBM, resulting in medium dimensionality and linear collisions within the quantum algorithm. 

\begin{figure*} 
    \centering
    \includegraphics[scale=0.95]{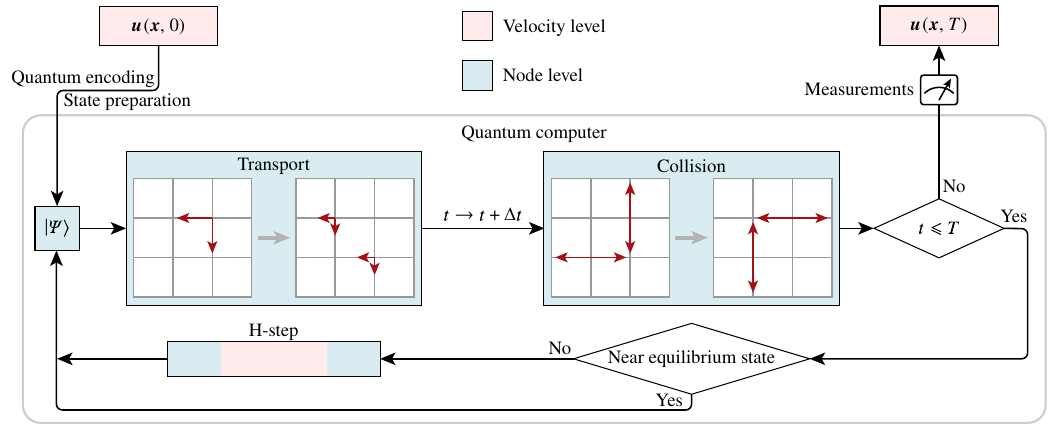}
    \caption{Flowchart for our QLBM. 
    The velocity field $\vec{u}$ is embed into the amplitudes of the quantum state $\ket{\varPsi}$ via probability encoding. 
    The preparation, measurement, and time evolution of the quantum state's density matrix is preformed on a quantum computer. When the state deviates significantly from equilibrium (quantified by the H-function), the H-step relaxes the ensemble back to equilibrium. During the H-step, the state density matrix alternates between velocity-level and node-level ensembles.  
    The state is measured to reconstruct the velocity field when the simulation is ended at a given time $t=T$.} 
    \label{fig:QLBM_flowchart}
\end{figure*}

The corresponding QLBM algorithm is sketched in Fig.~\ref{fig:QLBM_flowchart}. Initially, flow field is encoded into the quantum state $\ket{\varPsi}$. Then the quantum state evolves through transport and collision steps at the node level. 
When the state begins to deviate from equilibrium, an H-step is invoked to relax the state back to equilibrium. 
Upon reaching a specified time, the quantum state is measured to reconstruct the entire flow field. 

\vspace{.5cm}

\noindent\textbf{Ensemble description for QLBM}

\noindent The present QLBM employs the probability encoding method to embed the flow field into the amplitudes of quantum states within the lattice gas framework. 
We use the concept of the Gibbs ensemble \cite{Rivet_Boon_2001} to re-describe LBM and LGCA from the perspective of statistical mechanics at the velocity, node, and lattice levels, as illustrated in Fig.~\ref{fig:QLBM_levels}. 

The velocity-level ensemble, named after the velocity distribution, corresponds to classical LBM. 
In this ensemble, each realization within a node is illustrated by a ``card'' (containing a node) in Fig.~\ref{fig:QLBM_levels}(a). 
Each node is independent, and it involves $q$ independent realizations or cells \cite{wolf-gladrowLatticeGasCellular2000} with different velocity directions. 
Consequently, the dimensionality of this ensemble is $N_d = Nq$. 
Each realization at $\vec{x}$ and time $t$ occurs with the probability $p^{(n_j)}_j(\vec{x},t)$, where the subscript $j=1,\cdots,q$ denotes for a quantity along the $j$-th direction, and the Boolean variable $n_j(\vec{x},t)$ describes the occupation status of cells: $n_j=1$ indicates the presence of a particle with velocity $\vec{e}_j$, and $n_j=0$ indicates the absence of such a particle. 
The ensemble average of $n_j$ yields the velocity distribution function \cite{wolf-gladrowLatticeGasCellular2000} 
\begin{equation}
    f_j(\vec{x},t) = \sum_{n_j=0}^1 n_j p_j^{(n_j)} = p^{(1)}_j.
    \label{eq:f_and_p}
\end{equation}
On each card, the cells are color-coded by $p^{(1)}_j(\vec{x},t)$. 
Moreover, the fluid density $\rho$ and velocity $\vec{u}$ are defined as $\rho=\sum_{j=1}^q f_j$ and $\rho\vec{u}=\sum_{j=1}^q f_j\vec{e}_j$. 

The lattice-level ensemble, corresponding to LGCA, is a collection of all $N_d=2^{qN}$ possible states of the entire lattice, and all $N$ nodes in the lattice are fully correlated. Note that the excessively high $N_d$ in LGCA causes its quantum algorithm to lose speedup advantage. 
In Fig.~\ref{fig:QLBM_levels}(c), each realization for cellular automata is illustrated by a card with multiple nodes, and closed and open circles denote occupied and empty cells, respectively. 
At the lattice level, a realization is represented as $\bigotimes_{\mathcal{H}}\vec{n}(\vec{x})$, where $\vec{n}=\{n_1,\cdots,n_q\}$ denotes the state for each node, and $\bigotimes$ the Kronecker product of vectors over the Hilbert space $\mathcal{H}$ of node coordinates. 
Each card is color-coded by the probability $p_{\bigotimes_{\mathcal{H}}\vec{n}(\vec{x})}$ of each realization.

The node-level ensemble, between the velocity- and lattice-level ones, combines the low-dimensional representation of LBM with the linear collision treatment of LGCA.
In this ensemble, each node accommodates $2^q$ realizations $\vec{n}$, and nodes are independent of each other, unlike the lattice-level one where nodes are fully correlated. 
This key property leads to medium $N_d = 2^q N$. 
At the node level, we define the probability $p_{\vec{n}}(\vec{x})$ of $\vec{x}$, satisfying $\sum_{\vec{n}}p_{\vec{n}}=1$ within a node, and the joint probability $p_{\vec{x}, \vec{n}}$ of $\vec{x}$ and $\vec{n}$, satisfying $\sum_{\vec{x}, \vec{n}}p_{\vec{x}, \vec{n}} = 1$ over a lattice. 
Each card for realization $\{\vec{x}, \vec{n}\}$ is color-coded by $p_{\vec{x}, \vec{n}}$ in Fig.~\ref{fig:QLBM_levels}(b). 
The independence of nodes leads to $p_{\vec{x}, \vec{n}} = p_{\vec{n}} \rho/\sum_{\vec{x}} \rho$. 

\vspace{.5cm}
\noindent\textbf{Algorithm of QLBM}

\noindent The QLBM integrates certain components from both classical LBM and LGCA (see Methods). As illustrated in Fig.~\ref{fig:QLBM_flowchart}, all realizations $\{\vec{x},\vec{n}\}$ at the node level (blue panels) are encoded into a quantum state $\ket{\varPsi}$, and $\ket{\varPsi}$ is evolved in QLBM with ensemble transformation using a prediction-correction approach.

The prediction phase involves transport and collision steps, which are reformulated from those in LGCA to the node level. 
The transport step 
\begin{equation}
    \{\vec{x},\vec{n}\}\mapsto \sum_{j=1}^q \frac{n_j}{\rho_n} \{\vec{x}+\vec{e}_j,\vec{n}\}
    \label{eq:p_transport}
\end{equation}
is adapted from Eq.~\eqref{eq:substep_LGCA_trans}, where $\mapsto$ denotes a mapping and $\rho_n(\vec{x}, \vec{n}) \equiv \sum_{j=1}^q n_j$ denotes the number of particles at a node. 
In Eq.~\eqref{eq:p_transport}, the realization $\{\vec{x},\vec{n}\}$ is divided into $\rho_n$ components, and each component uniformly propagates into $\rho_n$ neighboring nodes with the equal proportion $n_j/\rho_n$. 

In the collision step, each possible collision event is encoded among the realizations in the ensemble, as illustrated by a collection of cards for a node in Fig.~\ref{fig:QLBM_levels}(b). This allows the LGCA collision model, originally developed at the lattice level, to be effectively used at the node level.  
Incorporating the collision step in Eq.~\eqref{eq:substep_LGCA_coll} into the node-level realization $\{\vec{x},\vec{n}\}$ yields the mapping $\{\vec{x},\vec{n}\}\mapsto  \{\vec{x},\vec{n}^\prime\}$ from the initial state $\vec{n}$ to the post-collision state $\vec{n}^\prime=\{n_1^\prime,\cdots,n_q^\prime\}$. Incorporating the collision probability $\gamma \in [0, 1]$ into this mapping leads to 
\begin{equation}
    \{\vec{x},\vec{n}\} \mapsto \gamma \{\vec{x},\vec{n}^\prime\} + (1-\gamma) \{\vec{x},\vec{n}\},     \label{eq:p_collision}
\end{equation}
where $\gamma$ governs the fluid viscosity. 
Note that the transport and collision steps are linear, positive, $L^1$-norm preserving, and numerically stable (see Secs.~\ref{sec:quantum_algorithm} and \ref{sec:stability} in SI). 
The linear operations can be efficiently represented as unitary transformations in quantum algorithms, and the positivity and $L^1$-norm preserving guarantee that the probability distribution remains physically realizable. 

During these linear transport and collision steps, the cells within a node can become correlated. This contradicts to the cell independence constraint at the velocity level, and causes the deviation of $f_j$ from the equilibrium velocity distribution $f_{\mathrm{eq}, j}$. 
This deviation, quantified by the H-function, results in unphysical results of QLBM (see Sec.~\ref{sec:H-theory} in SI). 
Thus, the correction phase in the QLBM algorithm introduces the H-step to keep $f_j$ near equilibrium, through inter-ensemble transformations between velocity and node levels across different $N_d$. 

The node-to-velocity-level transformation is through a linear and surjective mapping 
\begin{equation}
    p^{(n_j)}_j = \sum_{\substack{m=1 \\ m \ne j}}^q \sum_{n_m=0}^1 p_{\vec{n}} 
    \label{eq:f_node_relation}
\end{equation}
based on Eq.~\eqref{eq:f_and_p} and mass conservation. 
However, the inverse transformation is non-unique, because $N_d = N2^q$ at the node level is much higher than $N_d = Nq$ at the velocity level. 
Consistent with the independent cells within a node, we establish the velocity-to-node-level transformation 
\begin{equation}
    p_\vec{n} = p_{\mathrm{ind},\vec{n}} \equiv \prod_{j=1}^q p^{(n_j)}_j,
    \label{eq:independent_f}
\end{equation}
where the subscript ``$\mathrm{ind}$'' denotes the independence of individual $p^{(n_j)}_j$ components. 
This choice ensures that all possible $p_{\vec{n}}$ satisfy Eq.~\eqref{eq:f_node_relation}.
It is also consistent with the form of the local equilibrium probability  \cite{chenHtheoremGeneralizedSemidetailed1995} at a node, 
facilitating relaxing the ensemble back to equilibrium. 

\begin{figure}[ht!]
    \centering
    \includegraphics[scale = 1]{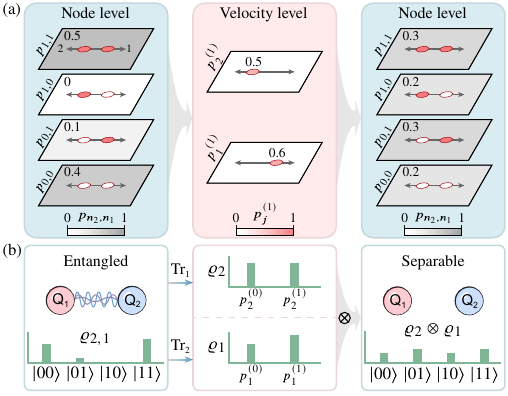} 
    \caption{(a) Schematic for inter-ensemble transformations in the H-step relaxing the ensemble toward equilibrium. 
    The color code and symbol meaning are the same as in Fig.~\ref{fig:QLBM_levels}. 
    A node-level ensemble (left panel), consisting of two cells and four realizations, can be transformed into a velocity-level ensemble (middle panel) with two realizations under mass conservation, where $p_1^{(1)}=p_{0, 1} + p_{1, 1}$ and $p_2^{(1)}=p_{1, 0} + p_{1, 1}$ are calculated by Eq.~\eqref{eq:f_node_relation}. 
    Assuming independence among cells, the velocity-level ensemble can be transformed back to node-level one (right panel), where $p_{0,0}=p_1^{(0)}p_2^{(0)}$, $p_{0,1}=p_1^{(1)}p_2^{(0)}$, $p_{1,0}=p_1^{(0)}p_2^{(1)}$, and $p_{1,1}=p_1^{(1)}p_2^{(1)}$ are calculated by Eq.~\eqref{eq:independent_f}.  
    Through these two transformations, the effective dimensionality is reduced from four to two, leading to smoothing of the probability distribution (right panel).  
    (b) Schematic for the quantum representation of the H-step. The histogram plots the probability distribution. 
    The two entangled qubits $Q_1$ and $Q_2$ (left panel), with density matrix $\varrho_{2,1}$ encoding $p_{n_2,n_1}$, can be reduced to a single qubit (middle panel) via the partial trace Tr$_1$ or Tr$_2$.
    Performing this operation on two identical $\varrho_{2,1}$ yields two independent qubits, with density matrices $\varrho_{1}$ and $\varrho_{2}$. 
    These two qubits form a separable state $\varrho_{2} \otimes \varrho_{1}$ (right panel).    
    In the H-step, the entangled qubits are disentangled, while the measurement outcomes of each qubit remain unchanged.
    }
    \label{fig:Hstep_schematic}
\end{figure}

As illustrated in Fig.~\ref{fig:Hstep_schematic}, the H-step applies Eqs.~\eqref{eq:f_node_relation} and \eqref{eq:independent_f}, relaxing distribution of $p_{\vec{n}}$ toward the local equilibrium one of $p_{\mathrm{ind},\vec{n}}$, along with the growth of quantum entropy (see Sec.~\ref{sec:H-theory} in SI). 
In the quantum representation, each cell at the node level is encoded by a single qubit. Entangled states correspond to cells that are not independent of each other, while separable states correspond to independent cells. 
The H-step is equivalent to disentangling the qubits without changing the measurement outcomes. 
This challenging quantum separability problem can hinder the overall speedup of the QLBM algorithm, and is expected to be resolved \cite{Bandyopadhyay_2009,Ogata_2006} in the future work.  

In the implementation of the QLBM algorithm, the probabilities are encoded into quantum states, and the states are evolved using quantum circuits (see Methods and Sec.~\ref{sec:quantum_algorithm} in SI).

\vspace{.5cm}
\noindent\textbf{Validation of QLBM}

\noindent We assess the proposed QLBM algorithm \cite{wolf-gladrowLatticeGasCellular2000} using three typical 2D incompressible flows, the Taylor-Green (TG) vortex, vortex pair merging, and decaying turbulence. In post-processing of simulation data, fluid density and velocity are calculated from $f_j$, and $f_j$ is calculated from Eqs.~\eqref{eq:f_and_p} and \eqref{eq:f_node_relation} where the probability is measured from the quantum state (see Sec.~\ref{sec:quantum_algorithm}E in SI). 

The QLBM simulations were conducted with the D2Q9 model (see Methods). Note that the NS equations can be derived from the D2Q9 model \cite{chen1989latticegasmodelwithtemperature}.  
It is straightforward to extend the QLBM to simulating 3D flows with more sophisticated DdQq models and more computational resources. 

\begin{figure}
    \centering
    \includegraphics[scale=0.95]{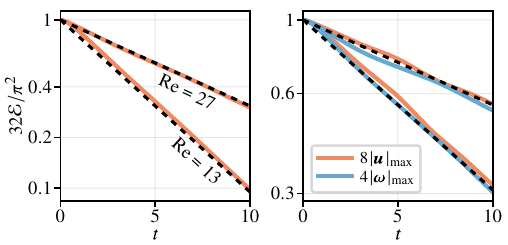}
    \caption{Comparison between QLBM results and the exact solutions for the 2D TG vortex at $\mathrm{Re} = 13$ and $27$. 
    Dashed lines represent normalized values of the exponentially decaying exact solutions for the total enstrophy $\mathcal{E} = \overline{\omega^2}/2 = \pi^2 \ee^{-4\nu t}/32$, for the maximum velocity magnitude $\abs{\vec{u}}_{\mathrm{max}} =  \ee^{-2\nu t}/8$, and for the maximum vorticity magnitude $\abs{\omega}_{\mathrm{max}}= \ee^{-2\nu t}/4$. 
    } 
    \label{fig:2DTG_vor}
\end{figure}

The QLBM of the TG vortex was conducted at $\mathrm{Re} = 13$ and 27 on $128^2$ and $256^2$ nodes, respectively, where the viscosity is calculated from Eq.~\eqref{eq:vis_LGA} with $\gamma=0.5$. 
This benchmark problem has the initial velocity $\vec{u}_0 = (\sin x \cos y,  - \cos x\sin y)/8$ and the exact solution of vorticity $\omega = \ee^{-2\nu t}\sin x \sin y/4$, showing the exponential decay of vortex intensity. 
The excellent agreement of the QLBM results and the exact solutions at different $\mathrm{Re}$ in Fig.~\ref{fig:2DTG_vor} indicates that the QLBM accurately captures viscous dissipation. 

\vspace{.5cm}
\noindent\textbf{QLBM simulation of vortex pair merging}

\begin{figure*}
    \centering
    \includegraphics[scale=0.9]{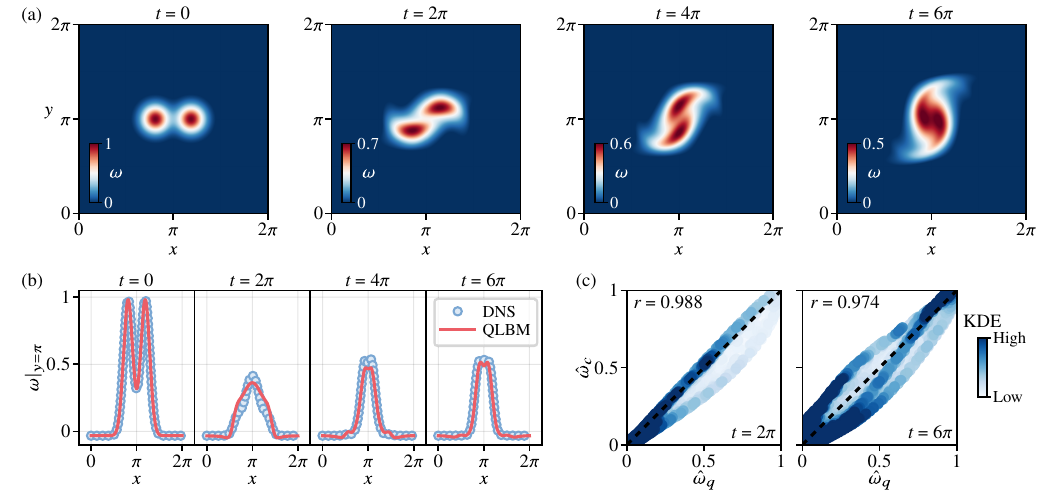}
    \caption{Comparison between the results of QLBM and DNS for the 2D vortex pair merging at $t=0$, $2\pi$, $4\pi$, and $6\pi$ with $\mathrm{Re} = 350$ on $2048^2$ nodes. 
    (a) Vorticity contours illustrate the nonlinear interaction of two merging vortices.
    (b) Cross-sectional vorticity profiles along $y=\pi$ from QLBM and DNS results.
    (c) Scatter plots of normalized vorticities $\hat{\omega}_q$ for QLBM and $\hat{\omega}_c$ for DNS, with correlation coefficients shown in the upper left. Data points are color-coded by the KDE.}
    \label{fig:merging_H_vor}
\end{figure*}

\noindent Merging of a pair of vortices involves nonlinear vortex interactions, which are crucial for many practical applications in vortical flows and turbulence. 
The initial vorticity is the superposition of the two Burgers vortices separated by a distance $l$. 
Each vortex is characterized by a Gaussian vorticity distribution $\omega=(\varGamma/2\pi\sigma^2) \exp(-r_v^2/2\sigma^2)$, with the radial distance $r_v$ from the vortex center. 
The QLBM of the vortex merging was conducted at $\mathrm{Re}=350$ on $2048^2$ nodes, with $\sigma = 0.1\pi$, $\varGamma = 0.2\pi$, and $l = 0.59$. %
Figure~\ref{fig:merging_H_vor}(a) illustrates the mutual rotation of two vortices driven by their induced velocities. 
The vortices first undergo persistent stretching and deformation, and then nonlinear vortex interaction and viscous diffusion initiate the merging process. 
Finally, the two vortices are merged into one with two spiral-like arms. 

The QLBM result is assessed using the result of the direct numerical simulation (DNS) for incompressible flow with the pseudo-spectral method \cite{xiong2019identifying,xie2019third}. 
Figure~\ref{fig:merging_H_vor}(b) demonstrates a good agreement between the vorticity distributions from QLBM and DNS, along the $y=\pi$ cross-section of the 2D vorticity field, where the migration of peaks in these distributions characterizes the vortex merging process. 
In general, our QLBM can accurately simulate nonlinear dynamics of vortex interactions.

Furthermore, the discrete vorticity from QLBM is transformed into a continuous probability distribution using the kernel density estimation (KDE). 
The correlation coefficient $r$ of normalized vorticities from QLBM and DNS decreases with time in Fig.~\ref{fig:merging_H_vor}(c). 
The KDE analysis reveals that slight discrepancies between QLBM and DNS in the low vorticity region. 
This discrepancy may stem from the deviation from equilibrium in the lattice gas system \cite{qianLatticeBGKModels1992}, which is attributed to the lack of growth of the H-function during the collision step.  
Here we selected $\gamma=1$ to achieve the largest $\mathrm{Re}$ (see Eq.~\eqref{eq:vis_LGA}) to clearly show the nonlinear vortex interaction, but this choice also causes a milder growth of the H-function during the collision step (see Sec.~\ref{sec:H-theory} in SI). 

\vspace{.5cm}
\noindent\textbf{QLBM simulation of decaying turbulence}

\noindent The quantum computing of turbulence is considered to be extremely challenging \cite{tenniequantum2025,fukagatatowards2022,Meng2023_Quantum}, 
because turbulence is highly nonlinear whereas quantum computing is based on linear operations.  
The present QLBM of turbulence showcases the capability of quantum computing for simulating one of the most challenging problems in classical physics \cite{Feynman2015_The}.   

The velocity of 2D decaying homogeneous isotropic turbulence is initialized using a Gaussian random field on $2048^2$ nodes with the spatially averaged initial velocity magnitude $\wb{|\vec{u}_0|}=0.03$ and a specified energy spectrum \cite{pope2001turbulent}, $E_k(k) \sim k^{-5 / 3} f_L(k )$ at $k<\sqrt{N}/3$, where $k$ is the wavenumber in Fourier space, and $f_L = (k/(k^2+4)^{1/2})^{5/3+2}$ characterizes the shape of $E_k$ in the energy-containing range. 
The maximum velocity magnitude $0.12\ll 1$ satisfies the low-Mach-number condition \cite{chen1998lattice}. 
Here $\mathrm{Re}=51$ is small, due to the small $\wb{|\vec{u}_0|}$ and the medium viscosity, which is proportional to $1/\sqrt{N}$ with modest $N$ and D2Q9 model \cite{succi2002colloquium}.  
To increase $\mathrm{Re}$, a larger $N$ and a more sophisticated DdQq model can be employed.  

\begin{figure*}
    \centering
    \includegraphics[scale=0.9]{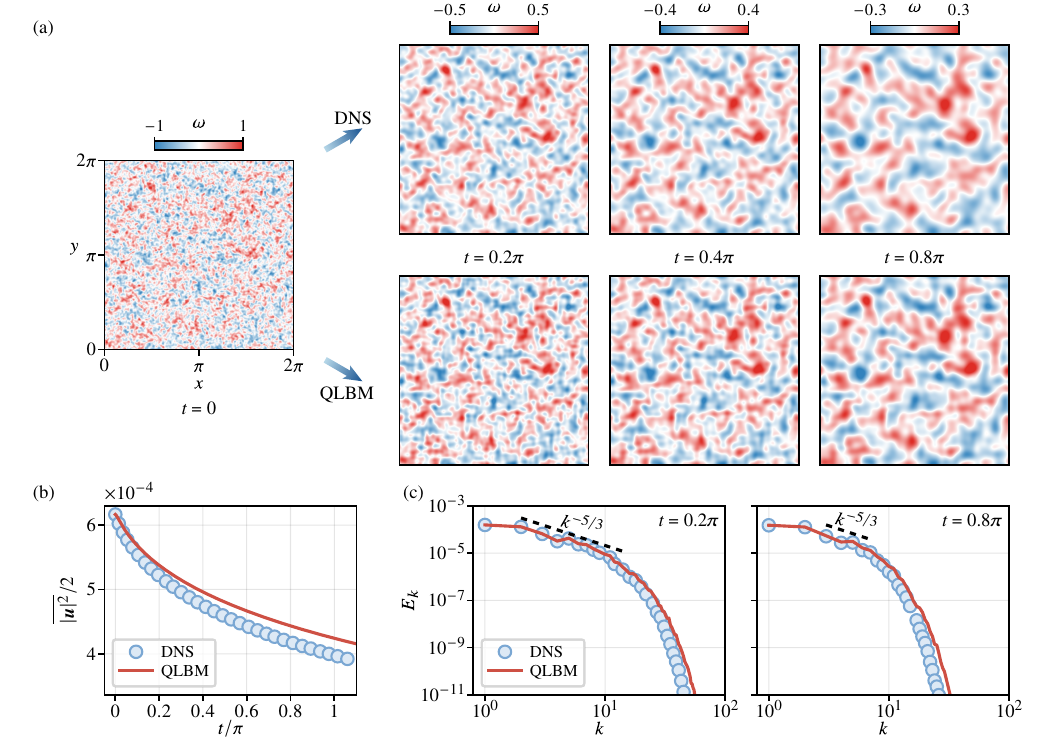}
    \caption{ 
    Comparison between the results of QLBM and DNS for 2D decaying homogeneous isotropic turbulence with $\mathrm{Re}=51$ on $2048^2$ nodes. 
    (a) Vorticity contours at $t=0.2\pi$, $0.4\pi$, and $0.8\pi$. (b) Evolution of the average energy. (c) Energy spectra at $t=0.2\pi$ and $0.8\pi$.
    }
    \label{fig:HIT}
\end{figure*}

The vorticity evolution in Fig.~\ref{fig:HIT}(a) exhibits consistent decaying rates and morphological characteristics between DNS and QLBM results, including the dissipation of small-scale flow structures and the generation of large-scale ones. 
In Fig.~\ref{fig:HIT}(b), the decay of the average kinetic energy $\wb{|\vec{u}|^2}/2$ in QLBM is slightly slower than that in DNS, perhaps due to the insufficient dissipation mechanisms at high wavenumbers in QLBM. 
The multiscale features and energy cascade are characterized using the energy spectrum in Fig.~\ref{fig:HIT}(c).
Consistent with the vorticity contour in Fig.~\ref{fig:HIT}(a), the low-wavenumber region of $E_k$ remains nearly constant over time, while the high-wavenumber region dissipates gradually, representing the viscous dissipation and inverse energy cascade at small scales \cite{lindborg1999can}. 
The slightly larger $E_k$ in QLBM than DNS at high wavenumbers indicates that QLBM has a slower energy transfer rate than DNS. 

\section*{Discussion}\label{sec:conclusions}
\noindent We propose the QLBM algorithm with ensemble transformations for simulating highly nonlinear fluid dynamics problems, including turbulence. The significant challenge in quantum computing for solving nonlinear problems is addressed by two conceptual steps. 
First, we transform a low-dimensional, nonlinear fluid system to a medium-dimensional, linear lattice gas system. The resultant node-level ensemble permits linear operations for particle transport and collision in a quantum algorithm. 
Second, the transformation between ensembles at different dimensions can cause inconsistent physical constraints, leading to the deviation of the velocity distribution from its equilibrium state. Thus we introduce the H-step to maintain the velocity distribution near equilibrium, which is equivalent to disentangling qubits without altering the measurement outcomes of each qubit.  

Our QLBM algorithm is linear, positive, and $L_1$-norm preserving. 
It can be implemented on a quantum computer without the need for classical computation, frequent intermediate measurements, or repetitive state preparation processes. Furthermore, it shows potential for exponential speedup in large-scale flow simulations (see Sec.~\ref{sec:complex} in SI). 

The QLBM has been validated by three typical 2D incompressible flows.  
The QLBM results show an excellent agreement with the exact solution for the linear TG vortex, 
and good agreements with the DNS results for the highly nonlinear vortex merging and decaying turbulence on $2048^2$ nodes. 
These successful applications of the QLBM showcase the capability of quantum computing to simulate highly nonlinear  dynamics. 

The QLBM incorporating ensemble transformation provides insights into quantum computing for nonlinear problems broadly and QCFD specifically. Integrating the understandings from both nonlinear classical systems and their quantum counterparts is essential for the development of an efficient quantum algorithm.

On the other hand, there are some issues to be addressed in the future work. First, the execution of the H-step demands significant qubit resources and may limit the potential speedup, and thus enhancing the efficiency of the H-step is essential. 
Second, the QLBM algorithm needs to be further improved for running on quantum devices, e.g., the current QLBM case requires $n_Q = 31$ qubits to encode the flow field and the quantum circuit depth can reach $n_Q^2$ for each time step, which poses a significant challenge for the current hardware \cite{bhartiNoisyIntermediatescaleQuantum2022}. 
Third, the QLBM is restricted to low Mach numbers, and has not incorporated non-slip boundary conditions. These issues will be addressed to enhance its applicability to more complex problems, such as multiphase and porous media flows.    

\section*{Methods}

\noindent\textbf{Recap of LBM and LGCA}

\noindent The Boltzmann equation can be discretized into the lattice Boltzmann equation $f_j\left(\vec{x}+\vec{e}_j \delta t, t+\delta t\right)-f_j(\vec{x}, t)=\Omega_j$ at the velocity level and the LGCA equation $n_j\left(\vec{x}+\vec{e}_j \delta t, t+\delta t\right)-n_j(\vec{x}, t)=\Omega_j$ at the lattice level, with the time step $\delta t$. 
The collision term $\Omega_j$ ensures the conservation of particles and momentum. 

Each time step in LBM can be decomposed into the transport step
\begin{equation}
        f_j(\vec{x},t)\mapsto f_j(\vec{x}+\vec{e}_j \delta t,t)\\
        \label{eq:substep_lbm_trans}
\end{equation}
and the collision step
\begin{equation}
    \begin{aligned}
        f_j(\vec{x}+\vec{e}_j \delta t,t)\mapsto & \ f_j(\vec{x}+\vec{e}_j \delta t,t)+\Omega_j(\vec{x}+\vec{e}_j \delta t,f_j)
        \\
        &=f_j\left(\vec{x}+\vec{e}_j \delta t, t+\delta t\right).
    \end{aligned}
\label{eq:substep_lbm_coll}
\end{equation}
Each time step in LGCA can also be divided into the transport step 
\begin{equation}
    n_j(\vec{x})\mapsto n_j(\vec{x}+\vec{e}_j \delta t)
    \label{eq:substep_LGCA_trans}
\end{equation}
and the collision step
\begin{equation}
    \begin{aligned}
        n_j(\vec{x}+\vec{e}_j \delta t,t)\mapsto & \ n_j(\vec{x}+\vec{e}_j \delta t,t)+\Omega_j(\vec{x}+\vec{e}_j \delta t,n_j)
        \\
        &=n_j\left(\vec{x}+\vec{e}_j \delta t, t+\delta t\right).
    \end{aligned}
\label{eq:substep_LGCA_coll}
\end{equation}

The collision in LBM is widely treated by the BGK model $\Omega_j = -\kappa (f_j - f_{\mathrm{eq},j})$ with the relaxation time coefficient $\kappa$ \cite{qianLatticeBGKModels1992}. 
Particularly, the local equilibrium velocity distribution $f_{\mathrm{eq}, j}$ is a nonlinear function of $\vec{u}$ and $\vec{e}_j$, prompting our attempt to adopt the linear collision model in LGCA. 
Under the collision rules in Fig.~\ref{fig:intro_HPP_D2Q9}, the collision event occurs with probability $\gamma$. 
We characterize the collision using the collision term in LGCA, which can be expressed as a linear combination of products of $n_j$ \cite{frischLatticeGasHydrodynamics1987}. 
This property makes it well-suited for implementation in quantum algorithms. 

\vspace{.5cm}
\noindent\textbf{Probability encoding for QLBM}

\noindent We encode the ensemble description of lattice gas at the node level into a quantum state $\ket{\varPsi}$ from a single qubit, through $q$ qubits, to $q + \log_2 N$ qubits, corresponding to encoding probabilities $p^{(n_j)}_{j}$, $p_{\vec{n}}(\vec{x})$, and $p_{\vec{x},\vec{n}}$, respectively.
First, we encode $p^{(n_j)}_j$ at the velocity level using a single-qubit state
\begin{equation}
    \label{eq:fi_qubit}
    \ket{f_j} \equiv \sqrt{p_{j}^{(0)}}\ket{0} + \sqrt{p_{j}^{(1)}}\ket{1}
\end{equation}
in a superposition of quantum basis states $\ket{0}$ and $\ket{1}$, where $\ket{0}$ and $\ket{1}$ encode $n_j=0$ and 1, respectively, and $p^{(0)}_j$ and $p^{(1)}_j$ are probabilities for measuring the qubit in states $\ket{0}$ and $\ket{1}$, respectively.  

\begin{figure*}
    \centering
    \includegraphics{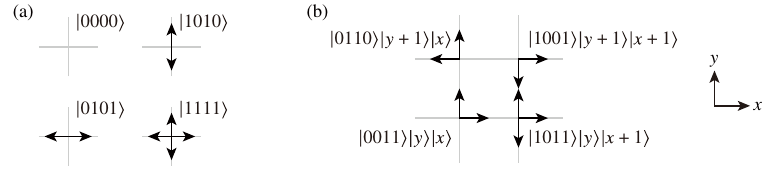}
    \caption{Encoding methods for the D2Q4 model. (a) Encoding of a node. For example, state $\ket{0101}$ for D2Q4 represents the presence of particles along $\vec{e}_1$ and $\vec{e}_3$. (b) Encoding of multiple nodes on a lattice. For example, consider a 2D field on a lattice with $N_x = 8$ and $N_y = 4$ nodes in the $x$- and $y$-directions, respectively. The coordinates for the sixth node along $x$ and the second one along $y$ are encoded into five qubits as $\vec{x} = (x^{(110)_2}, y^{(01)_2}) \leftrightarrow  \ket{\vec{x}} = \ket{11001}$, where $(\cdot)_2$ denotes the binary representation.}
    \label{fig:encode}
\end{figure*}

Second, we encode the node-level probability $p_{\vec{n}}$ in Eq.~\eqref{eq:independent_f} using $q$ qubits for each node. 
To encode all possible (up to $2^q$) states of a node, we employ a $q$-qubit pure state
\begin{equation}
    \ket{f_q \cdots f_1} \equiv \sum_{n_1, \cdots, n_q = 0}^{1} \sqrt{p_{\vec{n}}} \ket{n_q \cdots n_1},
    \label{eq:qubit_node}
\end{equation}
where $\ket{n_q \cdots n_1} = \bigotimes_{j=1}^q \ket{n_j}$ denotes basis states, and $n_1$ and $n_q$ are the lowest and highest indices of the qubit sequence, respectively.
As illustrated in Fig.~\ref{fig:encode}(a), we establish an encoding bijection $\vec{n} \leftrightarrow \ket{n_q \cdots n_1}$ between the node-level $\vec{n}$ and basis states $\ket{n_q \cdots n_1}$. 
The measurement of $\ket{n_q \cdots n_1}$ yields $p_{\vec{n}}$ in Eq.~\eqref{eq:qubit_node}, 
transforming the quantum state into the classical probability. 
Note that Eq.~\eqref{eq:qubit_node} can be decomposed into a product of $q$ separable qubits
\begin{equation}
    \ket{f_q \cdots f_1} = \bigotimes_{j=1}^q \ket{f_j}
    \label{eq:independent_qubit}
\end{equation}
at the velocity level, when $q$ cells for a node are independent. 
Consistent with Eq.~\eqref{eq:independent_f}, this encoding method can be transformed between the velocity and node levels when each $\ket{f_j}$ is independent.

Third, we encode the joint probability $p_{\vec{x},\vec{n}}$ for lattice $\mathcal{L}$. 
Besides encoding the node state using $q$ qubits, all coordinates $\vec{x} \in \mathcal{L}$ are encoded into the basis states by the binary representation $\vec{x} \leftrightarrow \ket{\vec{x}}$ using $\log_2 N$ qubits.  
As illustrated in Fig.~\ref{fig:encode}(b), each realization is encoded as $\{\vec{x}, \vec{n}\} \leftrightarrow \ket{n_q \cdots n_1} \ket{\vec{x}}$. 
Finally, we encode all realizations $\{\vec{x},\vec{n}\}$ at the node level into a $(q + \log_2 N)$-qubit quantum state
\begin{equation}
    \ket{\varPsi} \equiv \sum_{\vec{n}} \sum_{\vec{x}} \sqrt{p_{\vec{x},\vec{n}}} \ket{n_q \cdots n_1} \ket{\vec{x}}. 
    \label{eq:total_Hilbert}
\end{equation}

\vspace{.5cm}
\noindent\textbf{Quantum implementation of QLBM algorithm}

\noindent The QLBM algorithm is implemented using quantum circuits. 
As sketched in Fig.~\ref{fig:QLBM_flowchart}, the flow field is firstly transformed into the equilibrium velocity distribution $f_{\mathrm{eq}, j}$ based on macroscopic state, and then $f_{\mathrm{eq}, j}$ is encoded into the quantum state $\ket{\varPsi}$ at the node level using Eqs.~\eqref{eq:f_and_p}, \eqref{eq:independent_f}, and \eqref{eq:total_Hilbert}. Then, $\ket{\varPsi}$ sequentially undergoes the transport step, collision step, and H-step. 

In the quantum state preparation, the velocity and density fields are sequentially transformed into $f_{\mathrm{eq}, j}$, $p_{\vec{x},\vec{n}}$, and $\ket{\varPsi}$. During the transport step, qubits encoding coordinates undergo phase shifts in different directions, controlled by $\ket{f_j}$. Each basis is split into $\rho_{n}$ parts according to Eq.~\eqref{eq:p_transport}. 
In the collision step, all $\ket{f_j}$ exchange states according to predefined collision rules (see Fig.~\ref{fig:intro_HPP_D2Q9}). Collision occurs at a given probability $\gamma$, controlled by an ancilla qubit. 
In the H-step, $f_j$ is obtained by tracing off all qubits except the one encoding $\ket{f_j}$. 
The circuit is then repeated $q$ times to obtain all $f_j$, and all $f_j$ are combined by exchanging cross terms of different coordinates (see Sec.~\ref{sec:HStep_appendix} in SI). 
Finally, qubits are measured one by one to obtain $f_j$. 
The quantum circuit and implementation detail for each step are provided in Sec.~\ref{sec:quantum_algorithm} in SI. 

Note that the complexity of the H-step (see Sec.~\ref{sec:complex} in SI) could be a weakness of this algorithm. It causes linear growth in the number of qubits required for every H-step, and exponential growth in the initialization, transport, and collision steps. 
A possible improvement to the H-step lies in developing a more efficient and effective disentanglement algorithm  (see Sec.~\ref{sec:H-theory} in SI) that does not require repeating the circuit to prepare all $f_j$. 

\vspace{.5cm}
\noindent\textbf{Simulation setup of QLBM}

\noindent All QLBM simulations were performed in a periodic square domain with side $2\pi$, $N = N_x^2$ uniformly distributed nodes, and periodic boundary conditions.  
The time step is $\delta t = 2\pi/N_x$. 
The kinematic viscosity $\nu = \mu / \rho$ with $\rho = 1$ is determined by the viscosity
\begin{equation}
    \mu = \frac{e}{\gamma_{2D}e^2+\gamma_{2B}(1-e)^2}\delta t
    \label{eq:vis_LGA}
\end{equation}
with the internal energy $e$ and the subscript of $\gamma$ denoting the particle collision type (see Fig.~\ref{fig:intro_HPP_D2Q9}). 
We set $\gamma_{2D}=\gamma_{2B}=\gamma$ and $e=1/3$ \cite{chen1989latticegasmodelwithtemperature} in all simulations. 
The Reynolds number is defined by $\mathrm{Re} = 2\pi\wb{|\vec{u}_0|} / \nu$, where $\wb{(\cdot)}$ denotes the spatial average. 
The QLBM was implemented using our in-house code \cite{code}, which employs the quantum algorithm on a classical desktop computer. 
Only the diagonal elements of the density matrix for $\ket{\varPsi}$ were allocated, requiring the memory of $\mathcal{O}(2^{n_Q})$ bytes to realize the QLBM with up to $N_x^2 = 2048^2$ nodes.  

\section*{Data Availability}
\noindent The data presented in the figures and that support the other findings of this study will be publicly available upon its publication. 

\section*{Code Availability}
\noindent 
Simulation source codes have been deposited in QLBM (https://github.com/YYgroup/QLBM) \cite{code}. 

\section*{Acknowledgments}
\noindent This work has been supported by the National Natural Science Foundation of China (Grant Nos.~12432010, 11925201, and 11988102), the National Key R\&D Program of China (Grant No.~2023YFB4502600), and the New Cornerstone Science Foundation through the XPLORER Prize.

\section*{Author Contributions}
\noindent Y.Y. and B.W.~designed research. B.W. and Z. M.~performed research. All authors contributed to the analysis of data and writing of the manuscript.  

\bibliographystyle{modified-apsrev4-2.bst}
\providecommand{\noopsort}[1]{}\providecommand{\singleletter}[1]{#1}%

\let\addcontentsline\oldaddcontentsline

\clearpage
\onecolumngrid
\begin{center}
    \textbf{\large Supplementary Information for \\
    ``Quantum lattice Boltzmann method for simulating nonlinear fluid dynamics''}\\[.2cm]
    Boyuan Wang$^{1}$, Zhaoyuan Meng$^{1}$, Yaomin Zhao$^{1, 2}$ and  Yue Yang$^{1, 2, \dagger}$ \\[.1cm]
    {\itshape ${}^1$ State Key Laboratory for Turbulence and Complex Systems, \\ College of Engineering, Peking University, Beijing 100871, PR China} \\
    {\itshape ${}^2$ HEDPS-CAPT, Peking University, Beijing 100871, PR China} \\
    {\small $^\dagger$ yyg@pku.edu.cn}
\end{center}
\maketitle
\setcounter{page}{1}

\tableofcontents


\beginsupplement
\renewcommand{\thepage}{S\arabic{page}}
\renewcommand{\citenumfont}[1]{S#1}
\renewcommand{\bibnumfmt}[1]{[S#1]}

\section{Supplementary figures}

\begin{figure}[ht!]
    \centering
    \includegraphics{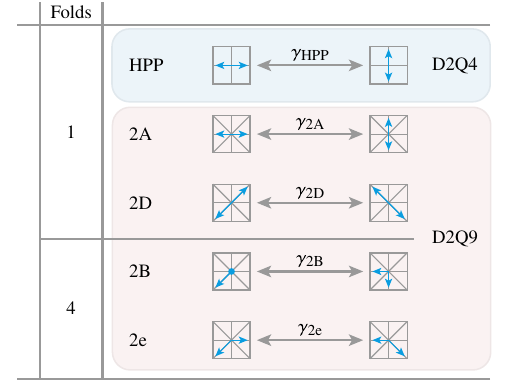}
    \caption{Two-body collision types in D2Q4 and D2Q9 models \cite{supp_chen1989latticegasmodelwithtemperature, supp_wolf-gladrowLatticeGasCellular2000} under the conservation of particles and momentum. 
    The first column (folds) represents the number of equivalent collision types. 
    The D2Q4 model only permits the head-on collision, and the D2Q9 model allows for more collision types.}
    \label{fig:intro_HPP_D2Q9}
\end{figure}

\section{Quantum circuits for QLBM algorithm}\label{sec:quantum_algorithm}
The QLBM algorithm is implemented using quantum circuits. 
As sketched in Fig.~\ref{fig:QLBM_flowchart}, the flow field is firstly transformed into $f_{\mathrm{eq}, j}$, and then $f_{\mathrm{eq}, j}$ is encoded into the quantum state $\ket{\varPsi}$ at the node level using Eqs.~\eqref{eq:f_and_p}, \eqref{eq:independent_f}, and \eqref{eq:total_Hilbert}. Then, $\ket{\varPsi}$ sequentially undergoes the transport step, collision step, and H-step. 
In our QLBM algorithm, both the transport and collision steps in Eqs.~\eqref{eq:p_collision} and \eqref{eq:p_transport} exhibit the crucial features for developing an efficient quantum algorithm as follows.
Linearity: the constant coefficients in Eqs.~\eqref{eq:p_collision} and \eqref{eq:p_transport} do not vary with $\vec{n}$ or $\vec{x}$.
Positivity: all coefficients, $\gamma \geq 0$ in Eq.~\eqref{eq:p_collision} and $n_j/\rho_n \geq 0$ in Eq.~\eqref{eq:p_transport}, are non-negative.
Preserving the $L^1$-norm $\sum_{\vec{x},\vec{n}}|p_{\vec{x},\vec{n}}|$: the initial one is set to be unity, and the sums of the coefficients in Eqs.~\eqref{eq:p_collision} and \eqref{eq:p_transport} remain unity. 

We will illustrate the quantum algorithm using the D2Q4 model \cite{supp_HPP1973}, owing to its mathematical simplicity. The generalization to other DdQq models is straightforward and will be discussed in Sec.~\ref{sec:different_models}. 
The D2Q4 model considers the 2D field $(\rho, u, v)$ for an incompressible, constant-density flow on a lattice consisting of $N = N_x \times N_y$ nodes. 
The particle velocity at each node can take $\vec{e}_1 = (1, 0)$, $\vec{e}_2 = (0, 1)$, $\vec{e}_3 = (-1, 0)$, or $\vec{e}_4 = (0, -1)$. The initial density $\rho=1$ is set at all nodes. 

\subsection{Quantum state preparation}\label{sec:Q_jnit}
In the quantum state preparation, the flow field $(\rho, u, v)$ is sequentially transformed into $f_{\mathrm{eq}, j}$, $p_{\vec{x},\vec{n}}$, and $\ket{\varPsi}$. 
First, we apply the Chapman-Enskog expansion \cite{supp_frischLatticeGasHydrodynamics1987} to truncate $f_{\mathrm{eq}, j}$ to terms of order $\mathcal{O}(|\vec{u}|^2)$. 
The truncated $f_{\mathrm{eq}, j}$ is set as the initial one $f_{0,j}$, where the subscript ``0'' denotes the initial state. 
Then, we transform $f_{0,j}$ into the velocity-level probabilities $p_{j}^{(1)}(x,y)=f_{0,j}(x,y)$ and $p_{j}^{(0)}(x,y)=1-f_{0,j}(x,y)$ 
using Eq.~\eqref{eq:f_and_p}. 
Second, under the assumption of independence in Eq.~\eqref{eq:independent_f}, 
the node-level probability for the D2Q4 model is $p_{n_1,n_2,n_3,n_4} = p_{1}^{(n_1)}p_{2}^{(n_2)}p_{3}^{(n_3)}p_{4}^{(n_4)}$. Note that, as time evolves, $n_1$, $n_2$, $n_3$, and $n_4$ in $p_{n_1,n_2,n_3,n_4}$ become dependent. This issue is further discussed in Sec.~\ref{sec:H-theory}. 
From Eq.~\eqref{eq:total_Hilbert} with $p_{\vec{x}, \vec{n}} = p_{\vec{n}}/A_{x,y}$ for satisfying the probability normalization condition and $A_{x,y} = N_x N_y$ for incompressible, constant-density flows, we obtain the initial quantum state $\ket{\varPsi_0} = \sum_{x,y} \sum_{n_1,n_2,n_3,n_4=0}^{1} \sqrt{p_{n_1,n_2,n_3,n_4}/A_{x,y}} \ket{n_4 n_3 n_2 n_1} \ket{x, y}$. 

\subsection{Transport step}\label{sec:Q_transport}

We implement the transport step in Eq.~\eqref{eq:p_transport} using a quantum circuit for the D2Q4 model. 
For example, the realization with $\vec{n} = \{0, 1, 0, 1\}$ ``transports'' along directions $\vec{e}_2$ and $\vec{e}_4$ to its neighbors at $(x,y+\delta y)$ and $(x,y-\delta y)$, respectively, i.e., $\{x,y,0,1,0,1\} \mapsto (\{x,y+\delta y,0,1,0,1\} + \{x,y-\delta y,0,1,0,1\})/2$. 

In the quantum presentation, we re-express Eq.~\eqref{eq:p_transport} into 
\begin{equation}
    \ket{n_4n_3n_2n_1}\ket{\vec{x}} \mapsto \frac{1}{\rho_n} \sum_{j=1}^4 n_j \ket{n_4n_3n_2n_1}\ket{\vec{x} + \vec{e}_j},
    \label{eq:basis_vec_trans}
\end{equation}
which evolves a basis state into a superposition of the states at neighboring nodes.
%
For clarity, we omit the notation $\ket{n_4n_3n_2n_1}$ in Eq.~\eqref{eq:basis_vec_trans}, as the qubits $n_1$, $n_2$, $n_3$, and $n_4$ act as control bits.
The transformation is then given by
\begin{equation}
    \ket{x, y} \mapsto \frac{1}{\rho_n} \left( n_1 \ket{x+1, y} + n_2 \ket{x, y+1} + n_3 \ket{x-1, y} + n_4 \ket{x, y-1} \right),
    \label{eq:basis_xy_trans}
\end{equation}
where each term represents the transport of particles to adjacent nodes with equal probability $1/\rho_n$.

We rewrite $\ket{\varPsi}$ in Eq.~\eqref{eq:total_Hilbert} into the density matrix
\begin{equation}
    \varrho=\ket{\varPsi}\bra{\varPsi}=\sum_{\vec{n},\hat{\vec{n}}} \ket{n_4n_3n_2n_1}\mathcal{D}_{\vec{n},\hat{\vec{n}}}\bra{\hat{n}_4 \hat{n}_3 \hat{n}_2 \hat{n}_1} =\begin{bmatrix}
        \mathcal{D}_{0000,0000}  & \cdots & \mathcal{D}_{0000,1111}\\
        \vdots  & \ddots & \vdots\\
        \mathcal{D}_{1111,0000}  & \cdots & \mathcal{D}_{1111,1111}\\
    \end{bmatrix}
    \label{eq:density_matrix}
\end{equation}
where $\mathcal{D}_{\vec{n},\hat{\vec{n}}}=\sum_{x,y} \sqrt{p_{\vec{n}}p_{\hat{\vec{n}}}{{A}}_{x,y}{{A}}_{x^\prime,y^\prime}}\ket{y, x}\bra{\hat{y}, \hat{x}}$ denotes the spatial distribution of $p_{\vec{n},\vec{x}}$ on the lattice. 
According to Eq.~\eqref{eq:basis_xy_trans}, Eq.~\eqref{eq:density_matrix} can be simplified in terms of  $\mathcal{D}_{\vec{n},\hat{\vec{n}}}$ into $\mathcal{D}\mapsto(n_1 \mathcal{T}(x+1) +n_2\mathcal{T}(y+1)+n_3\mathcal{T}(x-1)+n_4\mathcal{T}(y-1))\mathcal{D}/\rho_n$, as the quantum counterpart to Eq.~\eqref{eq:p_transport}. 
Here, the subscripts $\vec{n}$ and $\hat{\vec{n}}$ of $\mathcal{D}$ are omitted for clarity, and the state-shift operator $\mathcal{T}$ for transporting the quantum basis in one direction can be realized using efficient algorithms \cite{supp_shakeel_efficient_2020, supp_budinski_efficient_2023}. 

The state time evolution of $\mathcal{D}$ can be described by a trace-preserving completely positive (TP-CP) mapping \cite{supp_fujiwara_one--one_1999, supp_hayashi_quantum_2017}, $\mathcal{D} \mapsto \mathrm{Tr}_E \mathcal{U}^* \left(\mathcal{D} \otimes \varrho_E\right) \mathcal{U}$. 
Here, $\varrho_E$ denotes the environment density matrix of the environment vector $\ket{E}=(n_1\ket{00}+n_2\ket{01}+n_3\ket{10}+n_4\ket{11}) / \sqrt{\rho_n}$, 
where $\ket{00}$, $\ket{01}$, $\ket{10}$, and $\ket{11}$ encode directions $\vec{e}_1$, $\vec{e}_2$, $\vec{e}_3$, and $\vec{e}_4$, respectively; 
$\mathrm{Tr}_E$ denotes the partial trace over $\varrho_E$; 
$\mathcal{U}=\mathrm{Diag}\{ \mathcal{T}(x+1),\mathcal{T}(y+1),\mathcal{T}(x-1),\mathcal{T}(y-1)\}$ is a unitary operator on $\mathcal{D} \otimes \varrho_E$. 

\begin{figure*}
    \centering
    \includegraphics{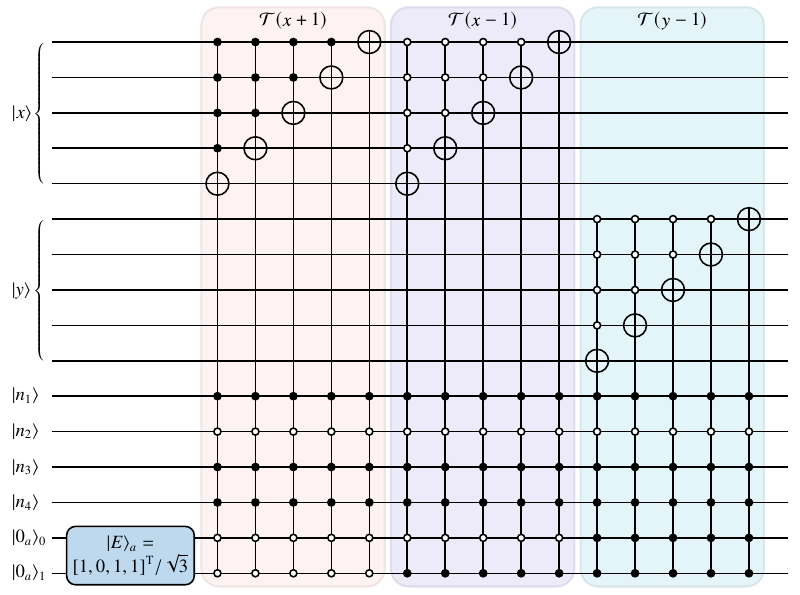}
    \caption{Quantum circuit of the transport step $\mathcal{D}\mapsto (\mathcal{T}(x+1)+\mathcal{T}(x-1)+\mathcal{T}(y-1))\mathcal{D}/3$ for the node state $\{n_1,n_2,n_3,n_4\}=\{1,0,1,1\}$ on $32\times 32$ nodes. This node state has three directions where particle exists with $\rho_n=3$. Two ancilla qubits are used to encode $\ket{E}$. }
    \label{fig:hpptransport}
\end{figure*}

The quantum circuit of $\mathcal{D}$ is illustrated in Fig.~\ref{fig:hpptransport}, where the block matrix is constructed using multiple multi-controlled X (MCX) gates. 
The full mapping of Eq.~\eqref{eq:basis_vec_trans} is achieved by extending $\mathcal{D}$ to encompass all $\vec{n}$.
Note that it is not necessary when $\rho_n \gg q/2$ at the Mach number $\mathrm{Ma} \ll 1$, due to $p_{\vec{x},\vec{n}}= \mathcal{O}(f_j^{\rho_n})$ with $f_j\ll 1$. 

\subsection{Collision step}\label{sec:Q_collision}
We implement the collision step in Eq.~\eqref{eq:p_collision} using a quantum circuit for the D2Q4 model. The collision term is given by
\begin{equation}
\Omega_{i}= n_{j+1} n_{j+3}\left(1-n_j\right)\left(1-n_{j+2}\right)-n_j n_{j+2}\left(1-n_{j+1}\right)\left(1-n_{j+3}\right),
\label{eq:coll}
\end{equation}
where $i=1,2,3,4$, $j = \mathrm{mod}(i,4)$. The two collision types, $n_{j+1} n_{j+3}(1-n_j)(1-n_{j+2})$ and $n_j n_{j+2}(1-n_{j+1})(1-n_{j+3})$, are depicted in Fig.~\ref{fig:intro_HPP_D2Q9}. 
As presented in Eq.~\eqref{eq:coll}, particles collide within a node when the particle distribution aligns with the specified collision types in Fig.~\ref{fig:intro_HPP_D2Q9}.  
Following the same procedure in Eqs.~\eqref{eq:basis_vec_trans} and \eqref{eq:basis_xy_trans} for the transport step, we re-express Eq.~\eqref{eq:p_collision} into the basis form
\begin{equation}
\begin{aligned}
    \ket{0101}\ket{y, x} &\mapsto \gamma\ket{1010}\ket{y, x}+(1-\gamma)\ket{0101}\ket{y, x},\\
    \ket{1010}\ket{y, x} &\mapsto \gamma\ket{0101}\ket{y, x}+(1-\gamma)\ket{1010}\ket{y, x},
\end{aligned}
    \label{eq:coll_hpp}
\end{equation}
where the coefficients of basis states $\ket{0101}\ket{y, x}$ and $\ket{1010}\ket{y, x}$ are exchanged with probability $\gamma$, while other basis states remain constant. 

\begin{figure}
    \centering
    \includegraphics{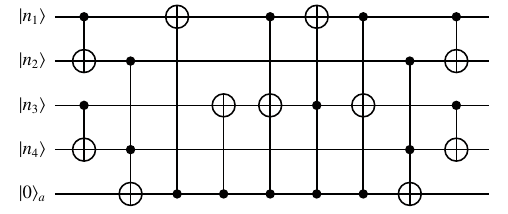}
    \caption{Quantum circuit of the collision step, where Eq.~\eqref{eq:coll_hpp} for $\gamma=1$ is realized using the method in Ref.~\cite{supp_loveQuantumExtensionsHydrodynamic2019} with one ancilla qubit. The exchange of basis states is characterized by the operator $\mathcal{T}_c:\ket{0101}\mapsto\ket{1010}$ and $\ket{1010}\mapsto\ket{0101}$, while leaves others unchanged.}
    \label{fig:hpp_coll_circuit}
\end{figure}

The logical form of Eq.~\eqref{eq:coll} is suitable for quantum encoding. 
For example, Eq.~\eqref{eq:coll} and the collision step in Eqs.~\eqref{eq:substep_LGCA_trans} and \eqref{eq:substep_LGCA_coll} can be re-expressed as $n_j^\prime = n_j \wedge \neg (n_j \wedge n_{j+2} \wedge \neg n_{j+1} \wedge \neg n_{j+3}) \vee (n_{j+1} \wedge n_{j+3} \wedge \neg n_j \wedge \neg n_{j+2})$, where the prime mark denotes the post-collision boolean state, and $\wedge$, $\vee$, and $\neg$ are logical operations AND, OR, and NOT, respectively. 
This logical form is used to implement Eq.~\eqref{eq:coll_hpp} on a quantum circuit, as illustrated in Fig.~\ref{fig:hpp_coll_circuit}. 
For arbitrary values of $\gamma$, the quantum circuit of Eq.~\eqref{eq:coll_hpp} can be realized using the TP-CP mapping $\varrho \mapsto \mathrm{Tr}_E \mathcal{U}^* \left(\varrho \otimes \varrho_E\right) \mathcal{U}$ with $\ket{E} =\sqrt{\gamma}\ket{0}+\sqrt{1-\gamma}\ket{1}$ involving a given value of $\gamma$, $\varrho_E=\ket{E}\bra{E}$, and $\mathcal{U}=\mathrm{Diag}\{ \on{I}, \mathcal{T}_c\}$, with the identity operator $\on{I}$ and the collision operator $\mathcal{T}_c:\ket{0101}\mapsto\ket{1010}$ and $\ket{1010}\mapsto\ket{0101}$. 

\subsection{H-step}\label{sec:Q_Hstep} 
\begin{figure}
    \centering
    \includegraphics{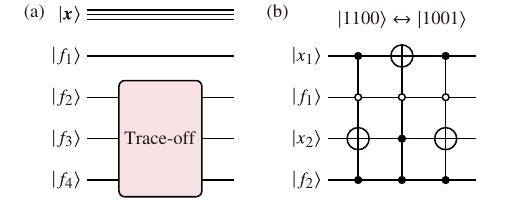}
    \caption{Quantum circuit of the part of H-step. (a) The process from lattice level to velocity level described in Eq.~\eqref{eq:f_node_relation}, with the output $f_1$. (b) The reverse process from velocity level to lattice level described in Eq.~\eqref{eq:Switch}, where only $f_1$ and $f_2$ with two coordinates are shown for an example.}
    \label{fig:H_traceoff}
\end{figure}

The distribution function $f_j$ can deviate from its equilibrium state $f_{\mathrm{eq}, j}$ in transport and collision steps. 
Meanwhile, the gas dynamics theory suggests that $f_j$ must be close to $f_{\mathrm{eq}, j}$, so that LBM or LGCA can be approximated by the NS equations \cite{supp_succi2002colloquium}. 
Thus, we introduce an additional H-step to keep $f_j$ near $f_{\mathrm{eq}, j}$ by re-setting $\ket{\varPsi}$ in a separable form as in Eq.~\eqref{eq:independent_qubit}, and quantify the deviation between $f_j$ and $f_{\mathrm{eq}, j}$ using the H-function (Gibbs entropy). 
The equilibrium state corresponds to the maximum H-function. 
In the implementation of transport and collision steps, insufficient growth of H-function can occur due to the entanglement among $\ket{f_j}$ (see Fig.~\ref{fig:Hstep_schematic} and Sec.~\ref{sec:H-theory}). 
The H-step comprises two stages given by Eqs.~\eqref{eq:f_node_relation} and \eqref{eq:independent_f}, which first transforms $p_{\vec{x},\vec{n}}$ from the node level (blue panels in Fig.~\ref{fig:QLBM_flowchart}) to $f_j$ at the velocity level (red panels in Fig.~\ref{fig:QLBM_flowchart}) and then transforms back to the node level. 
As elucidated in Sec.~\ref{sec:H-theory}, the H-step transforms a general $p_{\vec{n}}$ into the independence one $p_{\mathrm{ind},\vec{n}}$ defined in Eq.~\eqref{eq:independent_f}, corresponding to the growths of the H-function in classical statistical mechanics and the von Neumann entropy in quantum mechanics.  
The node-to-velocity-level mapping $\varrho\mapsto \varrho_j$, whose classical counterpart is Eq.~\eqref{eq:f_node_relation}, can be implemented by performing a partial trace 
\begin{equation}
    \varrho_j=\mathrm{Tr}_{1,2,3,4\neq j}\sum_{\vec{n},\hat{\vec{n}}} \sum_{x,y}\ket{n_4n_3n_2n_1}\bra{\hat{n}_4 \hat{n}_3 \hat{n}_2 \hat{n}_1}{\sqrt{p_{\vec{n}}p_{\hat{\vec{n}}}{{A}}_{x,y}{{A}}_{\hat{x},\hat{y}}}}\ket{y, x}\bra{\hat{y},\hat{x} }
    \label{eq:density_partial}
\end{equation}
of $\varrho$ in Eq.~\eqref{eq:density_matrix} over all indices except index $j$, as illustrated in Fig.~\ref{fig:H_traceoff}(a).  
Summing over the indices in Eq.~\eqref{eq:density_partial}, the diagonal elements of $\varrho_j$ become $\mathrm{Diag}[\varrho_j]= \{{p_{j}^{(0)}(x,y)}{{{A}}_{x,y}},{p_{j}^{(1)}(x,y)}{{{A}}_{x,y}}\}$, where $x$ and $y$ take all possible values on lattice $\mathcal{L}$ in sequence. 
%
Thus, $p_{\vec{n}}$ at the node level is transformed to $p_{j}^{(n_j)}$ at the velocity level,  
which is represented by the diagonal elements of $\varrho_j$ in Eq.~\eqref{eq:density_partial}, and is associated to $f_j$ in Eq.~\eqref{eq:f_and_p}. 
Note that the encoding of each $f_j$ requires $n_Q$ qubits, so the encoding of all $q$ distribution functions requires a total of $q n_Q$ qubits. 

On the other hand, implementing the velocity-to-node-level mapping $\varrho_j \mapsto \varrho$, whose classical counterpart is Eq.~\eqref{eq:independent_f}, presents greater complexity, because each $f_j(x,y)$ in $\varrho_j$ is associated with its individual coordinate. 
The goal of $\varrho_j \mapsto \varrho$ is to integrate all $f_j(x,y)$ in $\varrho_j$ from Eq.~\eqref{eq:density_partial} while preserving only a unified coordinate, as defined by $\varrho$ in Eq.~\eqref{eq:density_matrix}.
The derivation and implementation of $\varrho_j \mapsto \varrho$ is detailed in Sec.~\ref{sec:HStep_appendix}. 
Its quantum implementation relies on the state exchange circuit illustrated in Fig.~\ref{fig:H_traceoff}(b),  
by executing this quantum circuit for each qubit encoding $\vec{x}$ and performing a partial trace over the extra qubit. 

The H-step is only applied when the system significantly deviates from equilibrium, as illustrated in Fig.~\ref{fig:QLBM_flowchart}. 
The time interval $\delta t_H \ge \delta t$ for applying the H-step 
is primarily influenced by viscosity, since $\nu$ is proportional to $\delta t$ in Eq.~\eqref{eq:vis_LGA}.
In general, decreasing $\delta t_H$ increases simulation accuracy and cost. 
In all the present simulations, we selected $\delta t_H = \delta t$ to ensure the accuracy. 
The physical interpretation and validation of the H-step are elaborated in Sec.~\ref{sec:H-theory}. 

\subsection{Measurement}\label{sec:Q_measure} 
We extract the flow field from quantum state $\ket{\varPsi}$ through a series of measurements. 
Since $p_{\vec{x},\vec{n}}$ has been encoded along the diagonal elements of the matrix in Eq.~\eqref{eq:density_matrix}, we employ a projection-valued measure (PVM) to extract the field information from the coefficients of basis states. 
The PVM projects the density matrix $\rho$ onto a flow field. 
For the D2Q4 model, this projection can be represented by a set of positive semi-definite measurement operators $M_{x,y,n_1,n_2,n_3,n_4}$, where the subscripts $\{x, y, n_1, n_2, n_3, n_4\}$ correspond to the node-level description of the lattice gas. 
In accordance with the properties of PVM, each $M_{x,y,n_1,n_2,n_3,n_4}$ corresponds to the basis $\ket{n_4 n_3 n_2 n_1} \ket{y, x}$ and represents a rank-one projection Hermitian matrix.

According to the Born rule, the probability for obtaining $\{x,y,n_1,n_2,n_3,n_4\}$ in PVM is 
\begin{equation}
    p^{\on{M}}(x,y,n_1,n_2,n_3,n_4)=\mathrm{Tr}\big(\varrho M_{x,y,n_1,n_2,n_3,n_4}\big).
    \label{eq:PVM_condition_p}
\end{equation}
Substituting Eq.~\eqref{eq:density_matrix} and matrix elements $(M_{x,y,n_1,n_2,n_3,n_4})_{j, k} = \delta_{j, \iota}\delta_{\iota, k}$ with $\iota = (8n_4 + 4n_3 + 2n_2 + n_1)N + y_iN_x + x_i$, $x_i=0,1,\cdots,N_x-1$, $y_i = 0,1,\cdots,N_y-1$, and $y_iN_x+x_i=0,1,\cdots,N_xN_y-1$ into Eq.~\eqref{eq:PVM_condition_p} yields 
\begin{equation}
    p^{\on{M}}(x, y, n_1, n_2, n_3, n_4) = p_{x, y, n_1, n_2, n_3, n_4}.
    \label{eq:PVM_proof}
\end{equation}
Thus, we extract $p_{\vec{x},\vec{n}}$ at the node level by measuring all qubits with PVM. 

In the following post-processing to reconstruct the flow field, $p_{\vec{n}} = p_{\vec{x},\vec{n}} A_{\vec{x}}$ is calculated from $p_{\vec{x},\vec{n}}$ in Eq.~\eqref{eq:PVM_proof}. 
The distribution function $f_j$ is then calculated using Eqs.~\eqref{eq:f_node_relation} and \eqref{eq:f_and_p}. Finally, macroscopic quantities $\rho$ and $\vec{u}$ are obtained from $\rho=\sum_{j=1}^q f_j$ and $\rho\vec{u}=\sum_{j=1}^q f_j\vec{e}_j$.

\subsection{Implementation for the general LBM model}\label{sec:different_models}

Besides the implementation example for D2Q4, we outline the implementation of quantum algorithms for other DdQq models.
These models more sophisticated than D2Q4 can incorporate thermal macroscopic variables, additional collision configurations, and increased transport directions. 
For example, the D2Q9 model includes the internal energy $e$, which determines the viscosity and has four legitimate collision configurations \cite{supp_chen1989latticegasmodelwithtemperature}.  
The velocities of cell are $\vec{e}_1=(0,0)$, $\vec{e}_2=(1,0)$, $\vec{e}_3=(0,1)$, $\vec{e}_4=(-1,0)$, $\vec{e}_5=(0,-1)$, $\vec{e}_6=(1,1)$, $\vec{e}_7=(-1,1)$, $\vec{e}_8=(-1,-1)$, and $\vec{e}_9=(1,-1)$. Transport in the directions $\vec{e}_6$, $\vec{e}_7$, $\vec{e}_8$, and $\vec{e}_9$ requires the combination of $\mathcal{T}(x)$ and $\mathcal{T}(y)$, e.g., $\mathcal{T}(x-1)\mathcal{T}(y+1)$ for $\vec{e}_7$.

\section{Underlying physics of the H-step}\label{sec:H-theory}
This section elaborates the underlying physics of the H-step which enables the system to maintain proximity to the equilibrium state. 
First, we demonstrate that the H-step maintains the lattice gas system within a bounded distance from equilibrium, where the distance is quantified by the H-function. 
The H-function monotonically increases during particle collision \cite{supp_succi2002colloquium}, eventually reaching its maximum value as the system approaches equilibrium. 

The H-function
\begin{equation}
    H_v = -\sum_{j=1}^q \sum_{s=0}^1  p^{(s)}_j \ln \frac{p^{(s)}_j}{{{\lambda}}_j}
    \label{eq:H_velocity}
\end{equation}
is defined under the velocity-level probability description, where the coefficient ${{\lambda}}_j$ is determined to ensure isotropy of the fourth-order velocity tensor and Galilean invariance \cite{supp_qianLatticeBGKModels1992}. 
With Eq.~\eqref{eq:f_and_p}, Eq.~\eqref{eq:H_velocity} is reformulated into a convex function  \cite{supp_chenHtheoremGeneralizedSemidetailed1995}
\begin{equation}
    H_v = -\sum_{j=1}^q\left(f_j\ln\frac{f_j}{{{\lambda}}_j} + (1-f_j)\ln\frac{1-f_j}{{{\lambda}}_j}\right)
    \label{eq:Hvf}
\end{equation}
in terms of $f_j$, as required for the H-function. 
Under the incompressible assumption with $f_j \ll 1$, Eq.~\eqref{eq:Hvf} can be approximated as
\begin{equation}
    H_v = -\sum_{j=1}^q f_j \ln\frac{f_j}{{{\lambda}}_j}.
    \label{eq:Hv_simplify}
\end{equation}

\begin{figure*}
    \centering
    \includegraphics{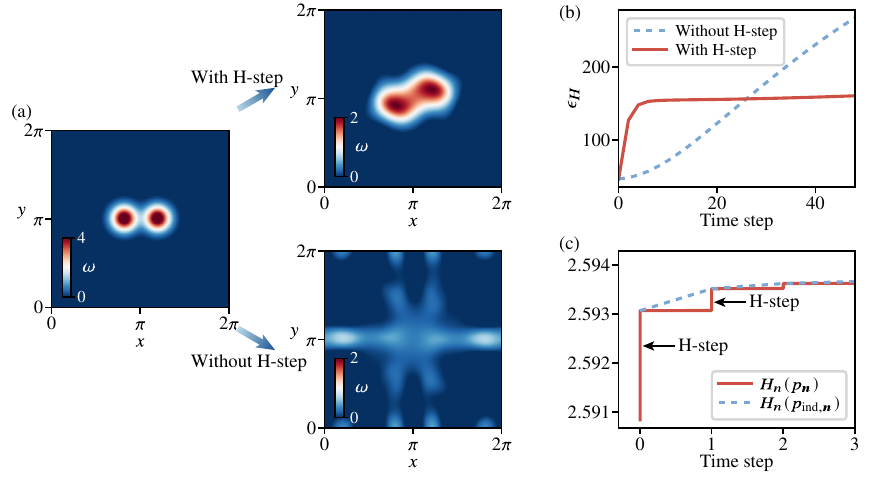}
    \caption{Effects of the H-step on QLBM of vortex pair merging (with $\nu = 0.015$ and $\delta t= 0.025$).
    (a) Vorticity contours in QLBM with and without the H-step at $t=\pi$. 
    (b) Deviation from the equilibrium state $\epsilon_H = H_{\mathrm{eq},v}-H_v$ versus time step. 
    (c) Evolution of $H_n(p_{\vec{n}})$ and $H_n(p_{\mathrm{ind},\vec{n}})$ versus time step. 
    }
    \label{fig:Hv}
\end{figure*}

The deviation from equilibrium during temporal evolution is quantified by $\epsilon_H = H_v(f_{\mathrm{eq},j})-H_v(f_j)$. Here, the equilibrium velocity distribution defined by
\begin{equation}
    f_{\mathrm{eq},j} = \rho\lambda_j\bigg(1+\frac{\vec{u}\cdot\vec{e}_j}{C_s^2}+\frac{(\vec{u}\cdot\vec{e}_j)^2}{2C_s^2}-\frac{\vec{u}\cdot\vec{u}}{2C_s^2}\bigg)
    \label{eq:equil_spe}
\end{equation}
is calculated from Eq.~\eqref{eq:Hv_simplify} by applying variational calculus \cite{supp_wagner1998h}, subject to the conservation laws, where $C_s = 1/\sqrt{3}$ represents the speed of sound. 
Figure~\ref{fig:Hv}(a) shows that the simulation without the H-step does not yield a physically reasonable result of vortex merging. 
Figure~\ref{fig:Hv}(b) demonstrates that $\epsilon_H$ remains approximately constant over time in the simulation with the H-step, indicating that the H-step constrains the system to maintain proximity to equilibrium. In contrast, the simulation without the H-step exhibits progressive deviation from equilibrium, ultimately leading to the non-physical result in Fig.~\ref{fig:Hv}(a). 

Next, we introduce the node-level H-function
\begin{equation}
    H_n = -\sum_{\vec{n}} p_{\vec{n}} \ln p_{\vec{n}}.
    \label{eq:H_node}
\end{equation}
It is more sensitive than $H_v$ in characterizing the process towards equilibrium in a lattice gas system, quantified by the monotonic growth of $H_n$, during the H-step. 
Note that $H_v$ grows during transport and collision steps, whereas it remains unchanged during the H-step, as $f_j$ is not altered in this step (see Eqs.~\eqref{eq:f_node_relation} and \eqref{eq:independent_f}). 
In contrast, $H_n$ can exhibit notable variation.
As shown in Fig.~\ref{fig:Hv}(c) for QLBM of vortex pair merging,  
$H_n(p_{\vec{n}})$ jumps at the application of the H-step and remains nearly constant during the transport and collision steps, suggesting that the H-step effectively increases $H_n$. 

The variation of $H_n$ reveals the underlying physics of the H-step. 
Similar to the velocity-level analysis, the equilibrium state with the local equilibrium probability $p_{\mathrm{eq}, \vec{n}}= \prod_{j=1}^q f_{\mathrm{eq}, j}^{n_j}(1-f_{\mathrm{eq}, j})^{n_j}$ within the node \cite{chenHtheoremGeneralizedSemidetailed1995} has the maximum $H_n$. However, calculating $p_{\mathrm{eq},\vec{n}}$ appears to be intractable in QLBM. As an alternative approach, we introduce an intermediate state $p_{\mathrm{ind},\vec{n}}$ in Eq.~\eqref{eq:independent_f} that satisfies the entropy condition \cite{supp_frischLatticeGasAutomataNavierStokes1986a, supp_frischLatticeGasHydrodynamics1987}
\begin{equation}
     H_n(p_{\mathrm{eq},\vec{n}}) \geq H_n(p_{\mathrm{ind},\vec{n}}) \geq H_n(p_{\vec{n}}),
     \label{eq:H_Frisch}
\end{equation}
which implies that the H-step transforms $p_{\vec{n}}$ into $p_{\mathrm{eq},\vec{n}}$, thereby increasing $H_n$ to approach its equilibrium value.  
As defined in Eq.~\eqref{eq:independent_f}, $p_{\mathrm{ind},\vec{n}}$ serves as an intermediate but computationally feasible estimate during the relaxation toward equilibrium. This theoretical result is validated by the numerical result in Fig.~\ref{fig:Hv}(c), where $H(p_{\mathrm{ind},\vec{n}})$ is consistently larger than $H(p_{\vec{n}})$ in the evolution. 

Second, we demonstrate the equivalence between the node-level H-function $H_n$ in the lattice gas system and the von Neumann entropy $H_{\mathrm{qc}} = -\mathrm{Tr}(\varrho \ln \varrho)$ in quantum information theory \cite{supp_bengtsson2017geometry}. 
Substituting the diagonal elements of the density matrix $\varrho$ from Eq.~\eqref{eq:density_matrix} into Eq.~\eqref{eq:H_node} yields
\begin{equation}
    H_{\mathrm{qc}} 
    = -\sum_{\vec{x}, \vec{n}} p_{\vec{n}}A_{\vec{x}}\ln(p_{\vec{n}}A_{\vec{x}})
    = -\sum_{\vec{n}}p_{\vec{n}}\ln p_{\vec{n}}\sum_{\vec{x}}A_{\vec{x}} -\sum_{\vec{x}}A_{\vec{x}}\ln A_{\vec{x}}
    = H_n - \sum_{\vec{x}}A_{\vec{x}}\ln A_{\vec{x}}.
    \label{eq:Hqc}
\end{equation}
By comparing Eqs.~\eqref{eq:Hqc} and \eqref{eq:H_node}, we obtain
\begin{equation}
    H_{\mathrm{qc}} = H_n,
    \label{eq:QC_CC_bridge}
\end{equation}
where the constant $\sum_{\vec{x}}A_{\vec{x}}\ln A_{\vec{x}}$ in Eq.~\eqref{eq:Hqc} has been omitted for clarity. 
Thus, the independence in classical systems and separability in quantum systems are intimately connected through entropy conditions. 
According to the additivity inequality of the von Neumann entropy \cite{supp_hayashi_quantum_2017}
\begin{equation}
    H_{\mathrm{qc}}\left( \varrho_{AB} \right) \leq H_{\mathrm{qc}}\left( \varrho_{A} \right) + H_{\mathrm{qc}}\left( \varrho_{B} \right)
    \label{eq:additivity}
\end{equation}
for any Hilbert spaces $\mathcal{H}_A$ and $\mathcal{H}_B$, where the subscripts $A$, $B$, and $AB$ denote system A, system B, and composite system of A and B, respectively. 
The equality condition holds if and only if systems A and B are separable, i.e., $\mathcal{H}_{AB} = \mathcal{H}_A \otimes \mathcal{H}_B$. 

Substituting $\ket{f_j}$ in Eq.~\eqref{eq:fi_qubit} to the right-hand side of Eq.~\eqref{eq:additivity} and applying Eq.~\eqref{eq:qubit_node} to the left-hand side yields the entropy condition
\begin{equation}
    H_{\mathrm{qc}}\left( \ket{f_q \cdots f_1} \right) \leq \sum_j H_{\mathrm{qc}}\left( \ket{f_j} \right).
    \label{eq:additivity_qubit}
\end{equation}
Its equality condition holds if and only if the qubits are independent as in Eq.~\eqref{eq:independent_qubit}. 
Therefore, Eq.~\eqref{eq:QC_CC_bridge} establishes a mathematical equivalence between the classical entropy condition in Eq.~\eqref{eq:H_Frisch} and quantum entropy condition in Eq.~\eqref{eq:additivity_qubit}. This implies that the entropy growth in the H-step is due to the disentanglement of each qubit, and it suggests the potential for applying quantum entropy properties to classical systems. 

\section{Numerical stability of QLBM}\label{sec:stability}

This section demonstrates that $\sum_\vec{x} H_n$ exhibits monotonic increase during both collision and transport steps, which guarantees the numerical stability of the QLBM \cite{supp_chen2000h}. 
First, the convexity of $H_n$ implies
\begin{equation}
    H_n\left(\frac{\sum_{\vec{n}}\xi(\vec{x},\vec{n})p_{\vec{x},\vec{n}}}{\sum_{\vec{n}}\xi(\vec{x},\vec{n})}\right)
    \ge 
    \frac{\sum_{\vec{n}}\xi(\vec{x},\vec{n})H_n(p_{\vec{x},\vec{n}})}{\sum_{\vec{n}}\xi(\vec{x},\vec{n})}
    \label{eq:convex}
\end{equation}
for non-negative coefficients $\xi(\vec{x}, \vec{n})$.
Substituting coefficients $\xi=\gamma$ for $\vec{n}^\prime$ and $\xi=1-\gamma$ for $\vec{n}$ from the collision model in Eq.~\eqref{eq:p_collision} into Eq.~\eqref{eq:convex} yields the relation 
\begin{equation}
        H_n\left(\gamma p_{\vec{x},\vec{n}^\prime} + (1-\gamma)  p_{\vec{x},\vec{n}}\right)\ge 
        \gamma H_n\left( p_{\vec{x},\vec{n}^\prime}\right) + (1-\gamma)H_n\left(p_{\vec{x},\vec{n}}\right)=H_n\left(p_{\vec{x},\vec{n}}\right),
        \label{eq:convex_coll}
\end{equation}
indicating that $H_n$ increases monotonically at any $\vec{x}$.
%
The equality in Eq.~\eqref{eq:convex_coll} holds when $\gamma$ equals either 0 or 1 at given $\vec{x}$.  
By summing Eq.~\eqref{eq:convex_coll} over all spatial coordinates $\vec{x}$, we demonstrate the monotonic increase of $\sum_\vec{x} H_n$ during the collision process. 
In particular, $\gamma=0.5$ for uniform mixing between $p_{\vec{x},\vec{n}^\prime}$ and $p_{\vec{x},\vec{n}}$ ensures a sufficient growth of $H_n$ in the collision step. 
 
Second, we prove the monotonic increase of $\sum_\vec{x} H_n$ during the transport step. Substituting Eq.~\eqref{eq:H_node} into $\sum_\vec{x} H_n$ yields
\begin{equation}
    \sum_\vec{x} H_n= \sum_{\vec{x},\vec{n}}\left(-p_{\vec{n}}(\vec{x})\ln p_{\vec{n}}(\vec{x})\right).
\end{equation}
We then utilize the convexity of entropy
\begin{equation}
    -p_{\vec{n}}\ln p_{\vec{n}} \ge -\sum_{j=1}^q \frac{n_j p_{\vec{n}}}{\rho_\vec{n}}\ln\left( \frac{n_j p_{\vec{n}}}{\rho_\vec{n}}\right)
    \label{eq:convex_transport}
\end{equation}
with decomposing $-p\ln p$ into $q$ distinct components.
By substituting the transport step in Eq.~\eqref{eq:p_transport} into Eq.~\eqref{eq:convex_transport}, we establish the relation
\begin{equation}
    -p_{\vec{n}}(\vec{x})\ln p_{\vec{n}}(\vec{x}) \ge -\sum_{j=1}^q \frac{n_j p_{\vec{n}}(\vec{x}+\vec{e}_j)}{\rho_\vec{n}}\ln  p_{\vec{n}}(\vec{x}+\vec{e}_j).
    \label{eq:convex_trans}
\end{equation}
Summing Eq.~\eqref{eq:convex_trans} over $\vec{x}$ and $\vec{n}$ yields
\begin{equation}
        \sum_\vec{x} H_n\left(\sum_{\vec{n}} \sum_{j=1}^q  \frac{n_j p_{\vec{n}}(\vec{x}+\vec{e}_j)}{\rho_\vec{n}}  \right)
        \ge
        \sum_\vec{x} H_n(p_{\vec{n}}(\vec{x})).
        \label{eq:transport_Hincrease}
\end{equation}
Its equality only holds for $j=1$, otherwise $\sum_\vec{x} H_n$ grows during the transport step. 
Thus, the transport step in Eq.~\eqref{eq:p_transport} generates entropy in QLBM, whereas it preserves entropy in classical LBM. 
This inherent entropy production contributes to enhancing numerical stability in QLBM.

\section{Implementation of the velocity-to-node-level mapping}\label{sec:HStep_appendix}

This section details the quantum implementation of the velocity-to-node-level mapping $\varrho_j \mapsto \varrho$, corresponding to Eq.~\eqref{eq:independent_f} in the H-step.  
Instead of using the density matrix, we employ a pure quantum state $\ket{\varPsi_j} = \sum_{x,y} \sqrt{A_{x,y}} \ket{f_j} \otimes \ket{y,x}$ for simplifying the mathematical representation. The diagonal elements of the state density matrix are identical to those in Eq.~\eqref{eq:density_partial} and equal to $p_j^{(n_j)}$. 

For the D2Q4 model, we represent the four states $\ket{\varPsi_j}$ encoding $f_1$, $f_2$, $f_3$, and $f_4$ as a composite system $\{\ket{\varPsi_1}, \ket{\varPsi_2}, \ket{\varPsi_3}, \ket{\varPsi_4}\}$. 
The state of the composite system is 
\begin{equation}
    \ket{\varPsi}=\bigotimes_{j=1}^4 \left(\sum_{x,y} \sqrt{{{A}}_{x, y}}\ket{f_j}\ket{y, x} \right).
    \label{eq:independent_HPP}
\end{equation}
There are four sets of coordinate bases $\ket{y,x}$ in $\{\ket{\varPsi_1}, \ket{\varPsi_2}, \ket{\varPsi_3}, \ket{\varPsi_4}\}$, whereas Eq.~\eqref{eq:independent_HPP} involves only a single set of coordinate bases.
Consequently, three of these sets are redundant.

We denote the four sets of coordinate bases as
\begin{equation}
    \mathcal{L}^{(4)} = \bigotimes_{i=1}^4 \mathcal{L},
    \label{eq:HPP_grid_4}
\end{equation}
which comprises $N^4$ basis states $\ket{y_1, x_1} \otimes \ket{y_2, x_2} \otimes \ket{y_3, x_3} \otimes \ket{y_4, x_4}$.
Each set of $\ket{y, x}$ corresponds to one $\ket{f_j}$ in Eq.~\eqref{eq:HPP_grid_4}. 
Among all bases in Eq.~\eqref{eq:HPP_grid_4}, $\ket{f_1}$, $\ket{f_2}$, $\ket{f_3}$, and $\ket{f_4}$ can only be correctly assembled when the conditions $x = x_1 = x_2 = x_3 = x_4$ and $y = y_1 = y_2 = y_3 = y_4$ are satisfied.

The basis states within $\mathcal{L}^{(4)}$ rarely satisfy this condition. 
To address this limitation, we propose a feasible approach to expand the space of legal states.
To elucidate this, we consider a 1D case with $N_x=2$, $q=2$, $x=0$ or 1, and two distribution functions $f_1$ and $f_2$. The state in Eq.~\eqref{eq:independent_HPP} is now simplified into a four-qubit mixed state
\begin{equation}\label{eq:psi2}
    \begin{aligned}
        \ket{\varPsi}
        &= \frac{1}{2}\left(\ket{f_2(0)}\ket{0} + \ket{f_2(1)}\ket{1} \right) \otimes \frac{1}{2}\left( \ket{f_1(0)}\ket{0} + \ket{f_1(1)}\ket{1}\right)
        \\
        &= \frac{1}{4}\left(\ket{f_2(0)}\ket{0} \ket{f_1(0)}\ket{0}+
        \ket{f_2(0)}\ket{0} \ket{f_1(1)}\ket{1}+
        \ket{f_2(1)}\ket{1} \ket{f_1(0)}\ket{0}+
        \ket{f_2(1)}\ket{1} \ket{f_1(1)}\ket{1}\right),
    \end{aligned}
\end{equation}
where $\ket{f_j(\cdot)}$ denotes the spatial position of $f_j$.

Our primary objective is to generate the entangled quantum state $\ket{f_2f_1}\ket{x}$ in Eq.~\eqref{eq:total_Hilbert}. The state transformation from the separable state $\ket{\varPsi}$ in Eq.~\eqref{eq:psi2} to the target entangled state 
\begin{equation}\label{eq:psi2_2}
    \ket{\varPsi^\prime}=  \frac{1}{2}\left(
    \ket{f_2(0)}\ket{+} \ket{f_1(0)}\ket{0}+
    \ket{f_2(1)}\ket{+} \ket{f_1(1)}\ket{1}\right)
\end{equation}
is formally defined by
\begin{equation}
    \ket{\varPsi}\mapsto\ket{\varPsi^\prime},
    \label{eq:Switch}
\end{equation}
where $\ket{+}=(\ket{0}+\ket{1})/\sqrt{2}$.
The transformation in Eq.~\eqref{eq:Switch} can be effectively implemented through basis state swapping
\begin{equation}
    \ket{1100} \leftrightarrow \ket{1001},
    \label{eq:fourqubit_switch}
\end{equation}
as implemented by the quantum circuit in Fig.~\ref{fig:H_traceoff}(b).

Equation~\eqref{eq:fourqubit_switch} modifies the cross terms $\ket{f_2(0)}\ket{0} \ket{f_1(1)}\ket{1}$ and $\ket{f_2(1)}\ket{1} \ket{f_1(0)}\ket{0}$ in Eq.~\eqref{eq:psi2}.
The cross terms can be written into
\begin{equation}
    \begin{aligned}
        \ket{f_2(0)}\ket{0} \ket{f_1(1)}\ket{1} =\ &  \sqrt{(1-p_1(0))(1-p_2(1))}\ket{0001}+\sqrt{p_2(0)(1-p_1(1))}\ket{1001}\\
        &+\sqrt{(1-p_2(0))p_1(1)}\ket{0011}+\sqrt{p_2(0)p_1(1)}\ket{1011},\\
        \ket{f_2(1)}\ket{1} \ket{f_1(0)}\ket{0} =\ &  \sqrt{(1-p_1(1))(1-p_2(0))}\ket{0100}+\sqrt{p_2(1)(1-p_1(0))}\ket{1100}\\
        &+\sqrt{(1-p_2(1))p_1(0)}\ket{0110}+\sqrt{p_1(1)p_2(0)}\ket{1110}.\\
    \end{aligned}
    \label{eq:cross_expand}
\end{equation}
Here, we denote $p_j(x)=p_j^{(1)}(x)$ for simplicity.
For the incompressible flow with $f_j \ll 1$ and consequently $p_1,p_2\ll 1$, neglecting the quadratic term $p_1p_2$ in Eq.~\eqref{eq:cross_expand} yields the first-order approximation %
\begin{equation}
    \begin{aligned}
        \ket{f_2(0)}\ket{0} \ket{f_1(1)}\ket{1} &=  \sqrt{1-p_1(0)-p_2(1)}\ket{0001}+\sqrt{p_2(0)}\ket{1001}+\sqrt{p_1(1)}\ket{0011},\\
        \ket{f_2(1)}\ket{1} \ket{f_1(0)}\ket{0} &=  \sqrt{1-p_1(1)-p_2(0)}\ket{0100}+\sqrt{p_2(1)}\ket{1100}+\sqrt{p_1(0)}\ket{0110}.\\
    \end{aligned}
    \label{eq:cross_expand2}
\end{equation}
Applying Eq.~\eqref{eq:fourqubit_switch} into Eq.~\eqref{eq:cross_expand2} yields 
\begin{equation}
    \begin{aligned}
        \ket{f_2(0)}\ket{0} \ket{f_1(1)}\ket{1} &=  \sqrt{1-p_1(0)-p_2(1)}\ket{0001}+\sqrt{p_2(0)}\ket{1100}+\sqrt{p_1(1)}\ket{0011},\\
        \ket{f_2(1)}\ket{1} \ket{f_1(0)}\ket{0} &=  \sqrt{1-p_1(1)-p_2(0)}\ket{0100}+\sqrt{p_2(1)}\ket{1001}+\sqrt{p_1(0)}\ket{0110}.\\
    \end{aligned}
    \label{eq:cross_expand3}
\end{equation}
By summing the two equations in Eq.~\eqref{eq:cross_expand3}, adding a quadratic term $p_1p_2$, and rearranging the terms under the assumption that $p_1, p_2 \ll 1$, we obtain
\begin{equation}
    \begin{aligned}
           \sqrt{1-p_1(1)-p_2(0)}\ket{0100}+\sqrt{p_2(0)}\ket{1100}+\sqrt{p_1(0)}\ket{0110}
           &\approx\ket{f_2(0)}\ket{1} \ket{f_1(1)}\ket{0}, \\
          \sqrt{1-p_1(0)-p_2(1)}\ket{0001}+\sqrt{p_2(1)}\ket{1001}+\sqrt{p_1(1)}\ket{0011}
          &\approx  \ket{f_2(1)}\ket{0} \ket{f_1(0)}\ket{1} .\\
    \end{aligned}
    \label{eq:cross_expand4}
\end{equation}
Substituting Eq.~\eqref{eq:cross_expand4} into Eq.~\eqref{eq:psi2} yields Eq.~\eqref{eq:psi2_2}, thereby accomplishing the transformation in Eq.~\eqref{eq:Switch}.

Then, by performing the partial trace over the third qubit in $\ket{\varPsi^\prime}$ in Eq.~\eqref{eq:psi2_2}, which encodes the extra coordinate, we obtain the mixed state 
\begin{equation}
   \frac{1}{2} \left( \ket{f_2(0)f_1(0)}\ket{0} + \ket{f_2(1)f_1(1)}\ket{1} \right),
   \label{eq:psi_4}
\end{equation}
which indicates that $\ket{f_1}$ and $\ket{f_2}$ are entangled through a shared coordinate system. 

The transformation in Eq.~\eqref{eq:Switch} can be implemented sequentially for each qubit encoding coordinate information. Through this iterative process, we obtain the entangled state $\ket{f_2f_1}\ket{x}$.
By performing three iterations of this procedure for D2Q4 (or $q - 1$ iterations for the general DdQq model), we combine all $f_j$ to construct $\varrho$, thereby implementing $\varrho_j \mapsto \varrho$. 

\section{Computational complexity}\label{sec:complex}

We analyze the computational complexity of the QLBM algorithm. 
In quantum computing, the algorithm complexity is primarily measured by the number of two-qubit gates required for implementation \cite{supp_Haferkamp2022}, especially in the NISQ era \cite{supp_preskill2018quantum, supp_bhartiNoisyIntermediatescaleQuantum2022}. 

The number of qubits required to encode an instantaneous flow field is $n_Q = \log_2 N + q$. 
The total qubit count for a simulation up to a given time $T$ is determined by $n_{Q,\mathrm{total}} = \left(q + N_H \right)n_Q$. It varies with the number $N_H= T/\delta t_H$ of H-steps, and each H-step implementation requires $q$ identical quantum circuits, as described in Sec.~\ref{sec:Q_Hstep}.

For the initial step, according to the Solovay-Kitaev (SK) theorem \cite{supp_Barenco_1995}, we initialize the quantum state in Eq.~\eqref{eq:total_Hilbert} into a quantum circuit using a unitary matrix. However, the gate complexity can reach up to $\mathcal{O}(4^{n_Q}/n_Q)$. 
To reduce the initialization cost, we only initialize $n_M$ dominant modes of the initial field while maintaining a controlled approximation error. 
For instance, the TG vortex has $n_M=1$ and the vortex merging has $n_M\approx 40$.
We approximate the two-norm distance to the exact initial density matrix as $\varepsilon\sim 1/(q n_M)$. 
Here, the trace of the density matrix is distributed across $q n_M$ degrees of freedom in a lattice gas system with $q$ cells, and each degree's amplitude takes the order of $1/(q n_M)$.
Based on the generalization of the SK theorem \cite{supp_dawson2005solovaykitaevalgorithm}, the gate complexity for circuit initialization is $\mathcal{O}(\log^c(1/\varepsilon))$ with $c<4$. 
Taking into account the complexity of the quantum Fourier transformation yields the gate complexity of the initial step $\mathcal{C}_{\mathrm{ini}} \sim 2 (1+\log_2 N)^2 + \log^c(q n_M)$. 

The gate complexity $\mathcal{C}_{\mathrm{trans}}$ of the transport step consisting of MCX gates is at the order of that for control qubits \cite{supp_Barenco_1995}. Summing the gate complexity of all MCX gates yields $\mathcal{C}_{\mathrm{trans}} \sim (q+\log_2q+\log_2N)\log_2Nq^2/2$. 
The gate complexity of the collision step, depending on the specific collision model and requiring only a negligible number of two-qubit gates, is $\mathcal{C}_{\mathrm{coll}} \sim q$. 
The gate complexity of the H-step $\mathcal{C}_{\mathrm{H}} \sim 3^2(q-1)\log_2N$ is illustrated in Fig.~\ref{fig:H_traceoff}(b). 
For each coordinate-encoded qubit, the process described in Eq.~\eqref{eq:Switch} requires three MCX gates, and $q-1$ operations to coherently link $q$ components of $f_j$.

To determine the total gate complexity of the circuit, we need to account for the numbers of transport steps, collision steps, and H-steps used in a QLBM simulation within a given time period $T$. 
Each implementation of an H-step requires $q$ pre-prepared quantum circuits encoding $f_j$. Consequently, for a quantum circuit undergoing $N_H$ H-steps, the simulation demands $q^{N_H-1}$ state initializations, $q^{N_H}$ transport and collision operations, and $q^{N_H-1}$ H-step computations. 
The total gate complexity can be expressed as $\mathcal{C}_{qc} =q^{N_H-1} (\mathcal{C}_{\mathrm{ini}}+\mathcal{C}_{\mathrm{H}})+q^{N_H}(\mathcal{C}_{\mathrm{trans}}+\mathcal{C}_{\mathrm{coll}})$. 
From the complexities for each part, we observe that $\mathcal{C}_{\mathrm{ini}}\sim \log_2^2 N$, $\mathcal{C}_{\mathrm{H}}\sim q\log_2 N$, and $\mathcal{C}_{\mathrm{coll}}\sim q$ become negligible compared to $\mathcal{C}_{\mathrm{trans}}\sim q^2\log_2^2N$, leading to the approximation $\mathcal{C}_{qc}\sim q^{N_H}\mathcal{C}_{\mathrm{trans}}$. 
By comparing $\mathcal{C}_{qc}$ with the classical computation cost $\mathcal{C}_{\mathrm{classic}}= Nq T/\delta t$ using the LBM, we obtain the ratio 
\begin{equation}
    \frac{\mathcal{C}_{qc}}{\mathcal{C}_{\mathrm{classic}}} \sim \frac{ q^{1+T/\delta t_H}  }{ T/\delta t}\frac{\log_2^2N}{N}.
    \label{eq:order_col}
\end{equation}

Equation \eqref{eq:order_col} demonstrates exponential speedup of QLBM over classical algorithms for flow simulation on a large-scale grid with $N>q^{N_H}$. 
However, this speedup criterion appears to be very demanding, e.g., $N>9^{50}$ for a QLBM simulation with the D2Q9 model and 50 H-steps. 
Note that the speedup depends on the choice of $\delta t \le \delta t_H \le T$ due to the term of $q^{1+T/\delta t_H}$ in Eq.~\eqref{eq:order_col}.  
Both the simulation accuracy and cost increase with decreasing $\delta t_H$. 
The total complexity of our QLBM is dominated by the H-step part, which requires repeated implementation of all other steps. 
Thus, enhancing the efficiency of the H-step, perhaps by using a more effective algorithm for qubit disentanglement, is essential to achieve exponential speedup in the QLBM. 

\bibliographystyle{modified-apsrev4-2.bst}
\providecommand{\noopsort}[1]{}\providecommand{\singleletter}[1]{#1}%

\end{document}